\documentclass[sigconf, nonacm]{acmart}

\usepackage{xcolor}

\usepackage{framed}
\usepackage{acmart-taps}

\usepackage{makecell}

\usepackage{subcaption}

\AtBeginDocument{%
  }

\copyrightyear{2026}
\acmYear{2026}
\setcopyright{cc}
\setcctype{by-nc-nd}
\acmConference[CHI '26]{Proceedings of the 2026 CHI Conference on Human Factors in Computing Systems}{April 13--17, 2026}{Barcelona, Spain}
\acmBooktitle{Proceedings of the 2026 CHI Conference on Human Factors in Computing Systems (CHI '26), April 13--17, 2026, Barcelona, Spain}
\acmDOI{10.1145/3772318.3791387}
\acmISBN{979-8-4007-2278-3/2026/04}

\begin{document}

\title{TermSight: Making Service Contracts Approachable}

\author{Ziheng Huang}
\orcid{0009-0001-8067-056X}
\affiliation{
\department{Siebel School of Computing and Data Science}\institution{University of Illinois Urbana-Champaign}
\city{Urbana}
\state{Illinois}
\country{USA}}
\email{zihengh2@illinois.edu}

\author{Tal August}
\orcid{0000-0001-6726-4009}
\affiliation{
\department{Siebel School of Computing and Data Science}\institution{University of Illinois Urbana-Champaign}
\city{Urbana}
\state{Illinois}
\country{USA}}
\email{taugust@illinois.edu}

\author{Hari Sundaram}
\orcid{0000-0003-3315-6055}
\affiliation{
\department{Siebel School of Computing and Data Science}\institution{University of Illinois Urbana-Champaign}
\city{Urbana}
\state{Illinois}
\country{USA}}
\email{hs1@illinois.edu}

\begin{abstract}
Legal contracts govern much of our society, but their specialized language is difficult for non-experts to read. While AI has enabled simplification of complex language, legal contracts pose unique challenges because of their connection to readers' values, ambiguity, and legally binding nature. Based on a formative study (N=20) using Terms of Service (ToS) as example contracts to study challenges in contract reading, we developed TermSight, an intelligent reading interface to probe the opportunities and challenges of designing augmentations for legal text. TermSight guides readers to relevant clauses with color-coded plain-language snippets of information and contextualizes ambiguous language with definitions and hypothetical scenarios. Importantly, TermSight's features always foreground the original, legally-binding contract text (e.g., linking to associated clauses). Our within-subjects study (N=20) demonstrated the opportunities of TermSight in making ToS significantly easier to read and navigate while revealing the challenges of augmenting service contracts such as ToS.
\end{abstract}




\maketitle

\begin{figure*}[!ht]
    \centering
    \includegraphics[width=0.825\textwidth]{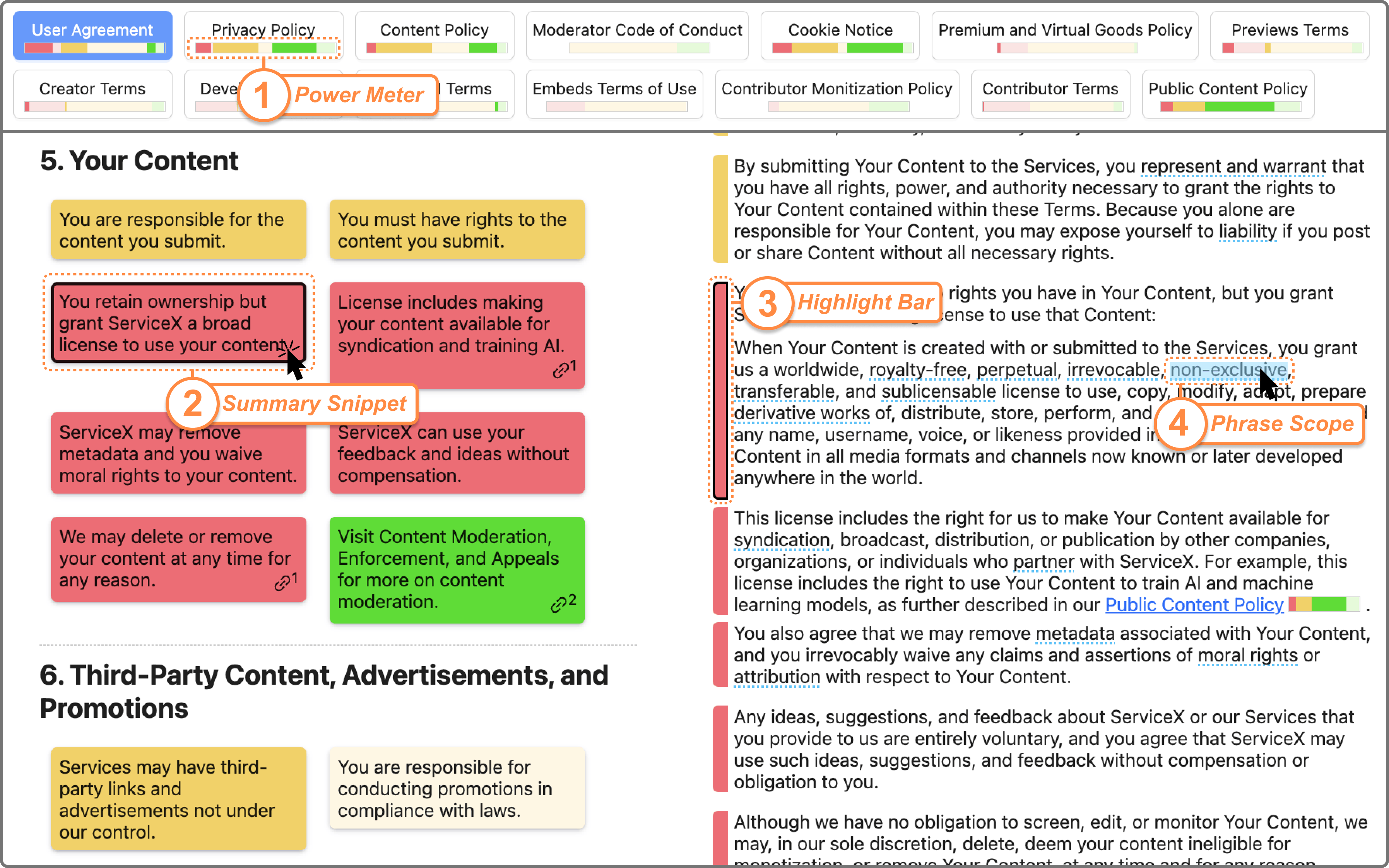}
    \caption{TermSight provides multi-level support for reading Terms of Service (ToS). 
    At the contract level, TermSight visualizes the relevance and power balance of content in each policy (1). 
    At the document level, TermSight chunks, summarizes, and categorizes content into one-sentence plain-language summaries (Summary Snippets) highlighted with colors that reflect power and relevance (2). 
    Readers can click the Summary Snippet (2) or Highlight Bar (3) to navigate between the two. 
    At the phrase level, TermSight offers phrase definitions and hypothetical scenarios for unfamiliar and ambiguous phrases (4).}
    \Description{The TermSight system provides guidance and reading support at multiple levels of Terms of Service (ToS) reading. The interface contains three panels: a top navigation panel for switching between different policies, the original text on the right, and summaries on the left. (1) At the contract level, TermSight visualizes the distribution of content in each policy in the top navigation panel as a horizontal bar plot based on its relevance to the user and power balance (favoring tenant vs. landlord). (2) Within each policy, TermSight chunks, summarizes, and categorizes content into one-sentence plain-language summaries (Summary Snippets), highlighted with colors that reflect power (red = favor provider, yellow = neutral, green = favor user) and relevance (low saturation = low relevance, high saturation = high relevance). The summaries are shown in the left panel, with corresponding highlight bars aligned to the original text on the right. (3) Readers can click either a Summary Snippet or its Highlight Bar to navigate between the summary and the source text. (4) At the phrase level, TermSight provides definitions and scenarios for unfamiliar or ambiguous terms.}
    \label{fig:termsight}
\end{figure*}

\section{Introduction}

Legal contracts govern much of our society. We sign contracts for where we live, who we work with, and, increasingly, for our digital interactions online. In order to encompass many possible situations and actions, contracts contain highly specialized language, such as: 
\begin{quote}
    \textit{``You grant us a worldwide, royalty-free, perpetual, irrevocable, non-exclusive, transferable, and sublicensable license to use, copy, modify, adapt, prepare derivative works of, distribute, store, perform, and display Your Content.''} --- Reddit Terms of Service ~\cite{reddittos}
\end{quote}
On the one hand, this specialized language allows contracts to articulate the entitlements, obligations, and prohibitions \citep{sancheti-etal-2023-read} of the signing parties. On the other hand, this specialized language is exceptionally difficult for many people without a legal background to read ~\cite{furth2017unexpected, mueller1970residential, amos2021privacy}. 

Recent AI systems, driven by language models (LMs), have expanded our ability to transform and augment text documents for different readers. Past reading systems have simplified scientific papers \citep{shin2024paper}, medical research \citep{august2023paper}, and news articles \citep{chen2023marvista} for general audience readers. These transformations have also begun to be applied to legal language, by, for example, generating summaries based on predefined categories of rights and responsibilities in contracts \citep{sancheti-etal-2023-read, pan2023toward} or through explaining legal concepts \citep{jiang2024leveraging}. 

However, legal text, and specifically contracts, pose unique challenges for simplification that current systems fail to address. First, contracts are inherently value-dependent: they are intended to encode and protect the values of the signing parties~\cite{haapio2017contracts,fiesler2016reality, sancheti-etal-2023-read}, which may lead to varying information needs among individuals.
Second, legal language is not just complex, but can be ambiguous (e.g., \textit{``We may share some of your personal information...''} ~\cite{bhatia2016theory}).
While this language provides the linguistic flexibility to manage unforeseen circumstances ~\cite{zhang2015elastic}, it also derails general audience readers ~\cite{ibdah2021should, tang2021defining,lebanoff2018automatic, bhatia2016theory, yerby2022deliberately}. Finally, the literal language of a contract is important: it is the only language that is legally binding ~\cite{cheong2024not}. 
Consequently, any text transformations should support rather than replace reading the original text.

In this paper, we explore augmentations that account for the unique challenges of simplifying legal contracts.
We focus on Terms of Service (ToS), the ubiquitous but rarely-read contracts that govern the use of digital services \citep{obar2020biggest, good2005stopping, ahmad2020policyqa}.
Despite consumers' concern for the information presented in ToS ~\cite{milne2004strategies, norberg2007privacy}, these contracts remain difficult to engage with. 
While prior work often focuses on a single policy within the ToS (e.g., a privacy policy ~\cite{nokhbeh2022privacycheck}), we explore challenges readers face when trying to engage with the full ToS contract.
In a formative study (N=20) where participants read ToS, we found that these contracts represented the extremes of the difficulties that legal contracts pose more generally (i.e., value-dependent, ambiguous, and legally binding), validating this initial focus. 
The parts of the ToS that were important to participants depended heavily on their envisioned usage and personal values, and this value-dependent information was lost among the many nested policies. ToS also contained complex and ambiguous terms that confused readers. Finally, existing features to support contract reading (e.g., policy names and section headers) often did not accurately represent the actual text and were not legally binding.\footnote{ToS often include clauses such as \textit{``Headings are used in these Terms for reference only and will not be considered when interpreting them''} ~\cite{reddittos}.}

In response, we designed a novel intelligent reading interface for ToS contracts, TermSight. 
In contrast to prior work that provides overview summaries detached from the original text ~\cite{TOSDR, privacyspy, pan2023toward}, we explore augmentations that support contract reading while acknowledging the unique characteristics of legal text. TermSight provides visual overviews of policies within a ToS (Power Meter, Figure~\ref{fig:termsight}.1), guides readers to relevant clauses with color-coded plain-language snippets of information (Summary Snippets, Figure~\ref{fig:termsight}.2), and contextualizes unfamiliar or ambiguous language with definitions and scenarios of potential implications (Phrase Scope, Figure~\ref{fig:termsight}.4). Importantly, TermSight's features always foreground and link to the original legal text.

To explore the opportunities and challenges of designing contract reading interfaces, we conducted a counterbalanced within-subjects study (N=20) using TermSight and a baseline HTML reader. 
Compared to prior focus on simplified reading tasks (e.g., reading a single clause~\cite{robinson2020beyond, lee2025could} or policy overview~\cite{tabassum2018increasing, korunovska2020challenges}), participants engaged with real ToS contracts with 10+ policies.
In the study, participants found reading ToS to be significantly easier with TermSight while also being more willing to read the original ToS. Participants reported that TermSight provided multi-level guidance and support while letting them drill down into clauses within the ToS important to them. 
We also observed that participants often used TermSight's generated text (e.g., Summary Snippets) as an entrance to, or a verification of, the original ToS text.
At the same time, our findings revealed the broader limitations of technological solutions like TermSight on facilitating a deeper understanding of ToS due to the overwhelming amount of potentially relevant information. 
We end by discussing the implications of our findings for inspiring future contract interfaces and regulatory innovation. In summary, this paper makes the following contributions:

\begin{enumerate}

    \item We characterize the challenges of contract reading, specifically ToS, including determining which policies to read, resolving ambiguous terms, and navigating misleading organization within the text. 
    
    \item We develop a novel intelligent reading interface, TermSight, to explore the design opportunities and challenges for helping readers engage with ToS while focusing on the original, legally binding language.

    \item We collect empirical evidence from our user study demonstrating the values of TermSight's features in finding relevant information, navigating ambiguous or complex language, and associating generated text with original text, while revealing the challenges of improving understanding of ToS.

\end{enumerate}

\section{Background and Related Work}

\subsection{Service Contracts}
A Terms of Service (ToS) is a type of standard form contract consisting of a body of policies and conditions outlined by a service provider, dictating the rules and expectations for both the service provider and the consumer. 
While prior work in HCI mainly focused on single policies within the ToS (e.g., a privacy policy ~\cite{nokhbeh2022privacycheck}), these contracts encompass a wide range of provisions such as copyright, privacy, returns, acceptable use, and service-specific terms ~\cite{amos2021privacy}. Important terms are often buried within this large body of documents ~\cite{srinath2020privacy}. ToS can vary in content~\cite{liu2021identifying, fiesler2016reality} and structure~\cite{ibdah2021should} across services. The language used to write ToS can also require advanced reading capability ~\cite{srinath2020privacy, amos2021privacy} and contain unfamiliar~\cite{ibdah2021should, tang2021defining} or ambiguous terminology~\cite{lebanoff2018automatic, bhatia2016theory, yerby2022deliberately}. Most consumers do not read modern ToS~\cite{obar2020biggest, bakos2014does}, despite the binding nature of ToS \citep{dickens2007finding} and consumers' self-reports indicating concern for the information presented in ToS ~\cite{milne2004strategies, norberg2007privacy}. This widespread disengagement highlights what Kar and Radin described as a ``paradigm slip'' in contract law, where agreements once premised on shared understanding and mutual consent are now reduced to unilateral boilerplate ``pseudo-contracts'', eroding fundamental principles of contract law and personal autonomy ~\cite{kar2019pseudo, palka2023terms, benoliel2019duty}. Our formative study further supplements prior findings by revealing the challenges readers face when navigating the entire ToS and the presence of misleading navigational affordances (\S\ref{sec:formative}).

\subsection{Augmenting Terms of Service}

Prior work has investigated new designs for readers navigating ToS ~\cite{kay2010textured,habib2021toggles,taber2020beyond}. \citet{kay2010textured} proposed Textured Agreements that use typographic manipulation, pull quotes, vignettes, and iconic symbols to improve reader attention and comprehension of privacy policy compared to plain text. \citet{habib2021toggles} investigated how icons and linked text could be designed to better convey privacy choices. \citet{taber2020beyond} proposed crowdsourced sentiment highlighting of sentences in ToS. Design recommendations on styling user agreements have also been proposed ~\cite{kay2010textured,habib2021toggles}.

Work has also explored helping consumers gain an overview of specific policies within a ToS, often the privacy policy (PP), without requiring reading the original document. 
Platform for Privacy Preferences (P3P) was an early attempt to standardize privacy policies by allowing service providers to submit privacy policies in a machine-readable format ~\cite{reagle1999platform}. Tools such as Privacy Bird build atop the P3P format to provide warnings when a site's PP does not match consumers' privacy preferences ~\cite{cranor2006user}. However, P3P was not widely adopted by service providers ~\cite{cranor2008p3p} and is no longer supported ~\cite{P3Pnotsupported}. 
Work has also proposed the use of a single-page summary for end-user license agreements (EULAs) ~\cite{good2005stopping, good2007noticing}, comic-based summaries for privacy policies ~\cite{tabassum2018increasing}, and Privacy Policy Nutrition Labels (PPNL) that label privacy policies in a table format ~\cite{kelley2009nutrition, reinhardt2021visual}. 

Other work has focused on providing overview summaries automatically or via crowdsourcing. ToS;DR uses volunteers to label and summarize terms in the privacy policy and the home page of ToS ~\cite{TOSDR}. Works like CLAUDETTE automatically detects a predefined set of potentially unfair clauses from ToS ~\cite{lippi2019claudette, guarino2021machine, braun2021nlp, braun2017satos}. PrivacyCheck~\cite{zaeem2018privacycheck, nokhbeh2020privacycheck, nokhbeh2022privacycheck}, PrivacyGuide~\cite{tesfay2018read}, and Polisis ~\cite{harkous2018polisis} extract terms relevant to a list of privacy-related questions and criteria and generate a grading for each. Past work has relied on either service providers or volunteers to make stylistic changes to contracts or automated tools to provide a summary external to the contract. In this paper, we explore new ways of combining these two threads to enable readers to navigate and read the original ToS made possible by the powerful text transformation capabilities of language models.

\subsection{Augmented Reading Interfaces}

Prior work has explored augmenting other types of documents, such as research papers \citep{10.1145/3659096} and news articles \citep{chen2023marvista}, with generative features. Work has provided faceted highlighting in research papers to support skimming ~\cite{fok2022scim} and context-sensitive definitions of terms and symbols ~\cite{head2021augmenting} to support reading. Also in research papers, Qlarify \citep{fok2024qlarify} enabled readers to expand paper abstracts with generative explanations. Systems have also provided automated question answering for research papers \citep{zhao2020talk} and medical papers \citep{august2023paper}. 
\citet{shin2024paper} transformed research papers into summaries of design implications. 
\citet{newman2024arxivdigestables} developed tools that synthesized multiple research papers to help researchers conduct literature reviews.
In the business domain, Marco allows readers to search and ask questions across collections of business documents ~\cite{fok2024supporting}. In the news domain, \citet{chen2023marvista} introduced Marvista, a reading interface that identifies the most summative portions of news articles.

In contrast to other documents explored previously, contracts---and specifically ToS---pose unique challenges to readers. Contracts directly connect with a reader's personal values: one reader might prefer stronger privacy safeguards while another might care about licenses ~\cite{haapio2017contracts,fiesler2016reality, sancheti-etal-2023-read}. Contract language is also meant to be interpreted; this means there are often ambiguous terms or incomplete information (e.g., the definition of `information' when discussing personal information collection), with the assumption that the reader has the requisite knowledge to fill in missing information ~\cite{bhatia2016theory}. Finally, contract language itself is the only legally binding text \citep{cheong2024not}, meaning any text transformations must maintain and display a tight connection with the original text. We take inspiration from past augmented reading systems to investigate what features can support readers in the unique context of contract reading.

\section{Formative Study}
\label{sec:formative}
We first establish the challenges general-audience readers face when reading a legal contract. We focus on Terms of Service (ToS) because ToS are legally binding contracts that are signed routinely in digital interactions. While ToS are rarely read completely \citep{obar2020biggest, good2005stopping, ahmad2020policyqa}, past legal work suggests that these contracts represent many of the broader issues faced in legal contracts today \citep{obar2020biggest, kar2019pseudo} and therefore present an ideal setting to explore possible reading features. Our formative study was guided by the following research questions:

\begin{quote}

\textbf{RQ1}: What information do readers want from a ToS? 

\textbf{RQ2}: What are the challenges readers encounter when trying to get this information from ToS?

\end{quote}

\subsection{Methodology}

\subsubsection{Study Procedure}
\label{sec:needfinding_method}

The entire study took 40 minutes, and participants were compensated 10\$. The study was approved by the Institutional Review Board (IRB) office at our organization. Participants first reported their past experiences interacting with ToS. Next, participants were asked to imagine as if they were a first-time user registering for a randomly assigned service and were provided 15 minutes to go through the ToS while speaking aloud any challenges. Participants were free to navigate to different policies within the ToS. 
After reading the ToS, we conducted semi-structured interviews about participants' experiences and challenges with ToS reading, along with opportunities for computational support. 
The interview questions are listed in Appendix \ref{Appendix_prelim}.

\subsubsection{Analysis}
\label{sec:needfinding_analysis}
We conducted a reflexive thematic analysis on the interview transcripts to identify common challenges and themes. We followed the six phases of reflexive thematic analysis suggested by Braun and Clarke ~\cite{braun2006using}. The lead researcher thoroughly explored the data, noted interesting features, systematically coded the data, and iteratively compared the codes to generate the initial themes. Through weekly meetings over the course of two months, the lead researcher discussed with other members of the research team to make sure the themes fit the data and further developed the themes. This included iterating on the descriptions of the themes and merging related themes.

\subsubsection{Materials}
\label{subsection:formative_materials}
Participants were randomly assigned one of the 10 services' ToS\footnote{Social Media: Reddit, Facebook, Instagram, Twitter; E-commerce: Amazon, eBay; Video: Youtube, Netflix; General: Google, Yahoo}. The 10 services were selected from the top 20 visited sites on semrush.com (2024/05), where we sampled the top 4 social media services, top 2 E-commerce services, top 2 video platforms, and top 2 internet services. We oversampled social media services because they were the most represented service category in the top 10 visited services.

\subsubsection{Participants}
We recruited 20 participants through Prolific (Male: 6, Female: 14). All participants were over 18, fluent in English, and located in the US. Participants' age ranged from 20 to 67 (\textit{\(\mu=35.75, \sigma=12.24 \)}). 3 participants had never read or skimmed a ToS, and 17 participants had read or skimmed at least 1 ToS before. We did not require participants to have prior experience in reading the ToS. All the participants had not read or did not remember reading the ToS they were assigned.

\subsection{RQ1: What information do readers want from a ToS? }

\subsubsection{\textbf{Participants have diverse, value-dependent information needs when reading ToS}}

All 20 participants described the need to know what control and rights the service had over the user, what the service was allowed to do, and how the user could be negatively impacted. 
While all participants shared a desire to know their general rights, different participants also noted more specific information needs, such as data collection and usage (11), intellectual property rights (8), purchasing and returns (7), account deletion by the service (5), arbitration and liabilities (3), content moderation (2), and exposure to offensive content or misinformation (1). 
10 participants explicitly mentioned how their information needs from the ToS and priorities depended on service usage and personal values. For example, P1 described that \textit{``I don't think any information is irrelevant. I just think some parts are more of a priority of knowing. If I happen to become a person who wants to create graphic designs, I would definitely utilize the Copyright portion of the ToS to ensure that I'm being compliant with [the platform's] expectations.''} On the other hand, P3, with a background in software, raised privacy concerns when using services for professional communication: \textit{``For Microsoft Teams and Gmail, I was using them for a collaborative project with a small company. I wanted to ensure that whatever data or information was shared would remain proprietary to us.''}

\subsection{RQ2: What are the challenges readers encounter?}
\label{sec:FormFindingsChallenges}

Our findings revealed that participants encountered challenges at the contract (\S\ref{sec:FormFindPolicy}), document (\S\ref{sec:FormFindDoc}), and phrase (\S\ref{sec:FormFindPhrase}) level. These challenges reflected the complexity and ambiguity of legal language. In addition, participants highlighted how existing navigational and reading affordances were often misleading and failed to support meaningful sensemaking of the actual ToS text (\S\ref{sec:FormFindDocMisl}).

\subsubsection{\textbf{Contract Level: Participants struggled to navigate nested policies and decide which policy to read.}}
\label{sec:FormFindPolicy}

Eleven participants mentioned that it's unclear to them what sub-policies are included when they are navigating the ToS because sub-policies are often hyperlinked throughout the documents. There often lacks a centralized list of what policies are included as part of ToS. Seven participants described the challenge of navigating across these nested policies as getting \textit{`lost in a whole tree of information'} (P18).
All 20 participants described relying on policy names to help decide whether to explore a policy. Yet, 17 participants noted that the names of the policies alone did not provide a clear mental model of what was in the sub-policies and whether there was important information they should know about. For example, P14 mentioned that: \textit{``When it's hyperlinked like `our rules and policies' and `privacy policy', it's difficult to determine if I need to know something from the link.''} 8 participants explicitly expressed their reluctance to click into the linked polices due to the lack of awareness of relevant information and the challenge of finding relevant information from a hyperlinked document.

\subsubsection{\textbf{Document Level: Participants lacked guidance and struggled to surface relevant information from extended, visually dense, and obfuscating text within a policy.}}
\label{sec:FormFindDoc}

Most (17/20) participants primarily skimmed the ToS. Six participants noted that they were not sure what they should look for in the ToS and might have unintentionally missed important information. P12 further commented how it's not always obvious if the information was relevant at first sight: \textit{``A lot of it seemed irrelevant until I started reading in a little bit more.''} Often, this inability to find information was connected to the overwhelming length of the document (18/20). Interestingly, many (13/20) participants also noted how text positioning seemed intentionally obfuscating: important information was positioned at the bottom of an extended policy. For example, P10 noticed how information about returns and cancellations of an E-commerce platform was positioned under the 13th section with the title `Additional Terms'. Similarly, three participants described how the use of friendly language tended to obfuscate information and made skimming more challenging. For example, P15 noticed language such as \textit{`We only use data to make [the service] a better place'} when reading a privacy policy. Yet, she later found out how user information can be used for targeted ads.\enlargethispage*{16pt}

\subsubsection{\textbf{Document Level: Participants found existing affordances to be ineffective and misleading}}
\label{sec:FormFindDocMisl}

Some of the ToS participants read contained navigational and reading affordances, such as interactive table of contents with section headers and overview summaries. However, 12 participants noted that, similar to policy names, section headings within policies were vague and not descriptive of the actual content. For example, when skimming the ToS, P10 skipped the `Content' section but later realized that the section was about intellectual property rights over user-generated content: \textit{``It just says `content' which is pretty vague when they're talking about getting exclusive intellectual rights. I feel like it's misleading. So people would skip right past it. I did the first time.''} In addition to ineffective navigational affordances such as vague policy names (17/20) and section headers (12/20), three participants accessed a sub-policy with a summary at the top. All three participants (P4, P14, P15) described using the summaries to gain an overview at the start. Yet, after reading the policy, participants noticed that the summary was missing key details: \textit{``The summary is kind of misleading. They gave you a more palatable version of [ToS]. If you scroll down, it will say we can terminate your account for no reason. They don't say that in the summary''} (P4).

\subsubsection{\textbf{Phrase Level: Participants struggled to contextualize unfamiliar or ambiguous terminology}}
\label{sec:FormFindPhrase}

When reading individual sections in a policy, 12 participants pointed out unfamiliar legal terminology (e.g., \textit{Arbitration}, \textit{Indemnity}, or \textit{class lawsuit}). These participants noted the difficulty to interpret the meaning of legal jargon in the context of their situation: \textit{``I can recognize these are legal terms, but I don't necessarily know what that means for me''} (P10). In a similar vein, eight participants noticed ambiguous terminology that made it challenging for them to understand the implications of signing (e.g., `we share your data with 3rd parties', and `retain certain information about you'). P15 mentioned that \textit{``I really think the big difficulty is the vagueness of the language and the constancy of exceptions that are vague, which gives them a lot more leeway''}.

\subsection{Design Guidelines}

We propose four design guidelines based on our observations that readers faced interrelated barriers at all levels of ToS reading:

\begin{quote}
\textbf{DG1}: Contract level: Help readers form mental models of each policy of a ToS to determine important documents to interrogate.

\textbf{DG2}: Document level: Help readers identify potentially relevant information within a policy. 

\textbf{DG3}: Phrase level: Help readers read and interpret original text with technical language, jargon, and ambiguous phrases.

\textbf{DG4}: All summaries should be tightly coupled with the text used to generate them. When reading any generated overviews, readers should be able to retrieve the original document text in at most one click. 
\end{quote}

\section{The TermSight System}

We reify these design guidelines in TermSight (Figure ~\ref{fig:termsight}), an augmented reading interface for exploring opportunities to make ToS contracts more approachable with features at every granularity:

~\begin{quote}

     \textbf{Power Meter:} A \textit{contract-level} visualization of the relevance of information within a document and the distribution of power between the user and the platform that information represents. 

    \textbf{Summary Snippets:} One-sentence plain language summaries of information chunks at the \textit{document-level}. Snippets are rendered with color and saturation that represent power and relevance to the user, similar to the Power Meter. 

    \textbf{Phrase Scope:} In-situ \textit{phrase-level} definitions and scenarios via a tooltip that contextualizes the meaning of the phrase while allowing users to ask clarification questions.
    
\end{quote}

We adopted an iterative design approach in developing TermSight. Eight participants evaluated an early prototype of TermSight, and their feedback informed the final design. 
Additional details about the iterative design study, results, and design changes can be found in Appendix ~\ref{appendix_iterativedesign}.

\subsection{Design Abstractions}
\label{sec:DesignAbstraction}

The design of TermSight is grounded in two design abstractions: 1) \textit{Information Snippets}, and 2) \textit{Power} and \textit{Relevance} classification.

\subsubsection{Information Snippets} 
\label{sec:infoSnippets}

Our formative study revealed that the information readers cared about in ToS can be buried within visually dense text. To enable sense-making of finer-grained pieces of information as opposed to an entire section, we use \textbf{Information Snippets} as the unit of interaction in TermSight.
An Information Snippet is a continuous span of original text that shares the same topic and can be summarized in a single sentence of up to 12 words (detailed in Section \ref{sec:infosummarySnippets}). These Information Snippets form the basis of the Summary Snippets (\S\ref{sec:summarysnippets}). Information Snippets are non-overlapping and together reconstruct the original document's content. Figure ~\ref{fig:summarysnippets} illustrates the process of how TermSight divides a paragraph into five Information Snippets, each paired with its Summary Snippet. 
The idea of designing interactions and sense-making around finer-grained Information Snippets was inspired by prior works that modularize and `objectify' tools~\cite{bier1993toolglass, bederson1996local},  attributes~\cite{xia2016object}, and AI agent's memory~\cite{huang2023memory} to enable interaction specificity and direct manipulation.

\subsubsection{Power and Relevance Classification}
\label{sec:powerrelevance}

\begin{figure}[!h]
    \centering
    \includegraphics[width=0.47\textwidth]{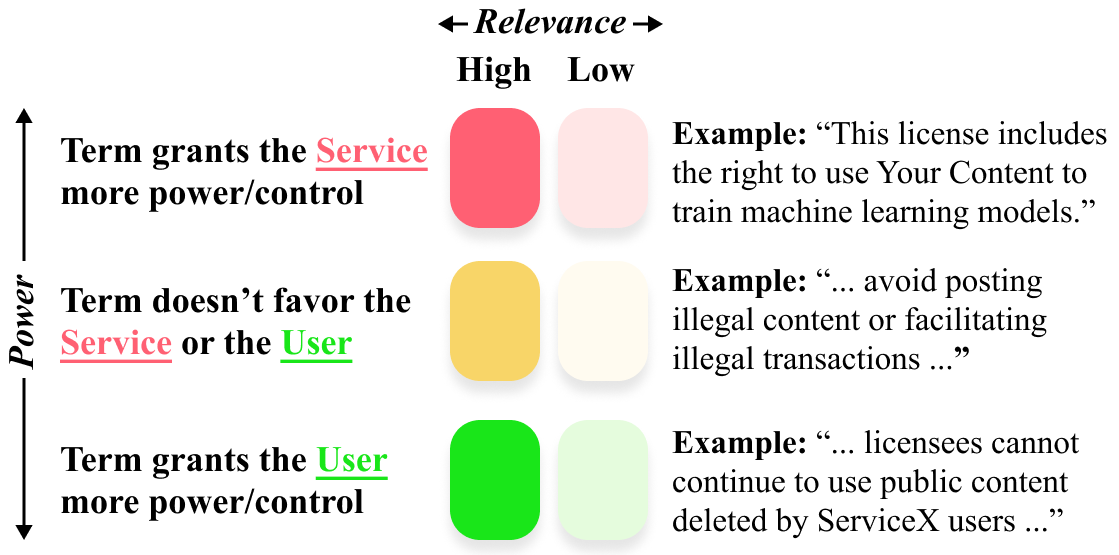}
    \caption{Design conceptualization of Power and Relevance. Power refers to the degree to which a snippet grants control to the service provider or the user. Relevance refers to whether or not the snippet is relevant to the user's persona.}
    \label{fig:colorscheme}
    \Description{The figure shows how TermSight defines and visualizes power and relevance using color. Power indicates the degree to which a snippet grants control to the service provider or the user (red = favor provider, yellow = neutral, green = favor user). Relevance indicates whether a snippet is pertinent to the user’s persona, which is based on intended service usage and personal values (low saturation = low relevance, high saturation = high relevance). The figure also provides examples. A provider-favoring clause: `This license includes the right for us to use your content to train machine learning models…’; A neutral clause: `Avoid posting illegal content’; and a user-favoring clause: `Our licensees cannot display public content deleted by the users…’.}
\end{figure}

Our formative study revealed participants’ information needs regarding the rights and control held by the service and the user. These needs may also vary depending on the intended service usage. To provide information scent ~\cite{pirolli1999information} for snippets of information that a reader may care about, we reify participants' information needs by defining
\textbf{Power} and \textbf{Relevance} as two qualities of an information---and thus a summary---snippet. Power is the degree to which a snippet's text grants control to the Service Provider or the User (Categories: Service, Neutral, or User). Relevance is whether the snippet's text is directly relevant to the user's intended usage of the service or their values (Categories: High, Low). 
For example, a snippet related to selling would have low relevance for a user more focused on buying.
As shown in Figure ~\ref{fig:colorscheme}, we visually encode Power and Relevance in TermSight with a combination of hue (Red, Yellow, Green) and saturation (High, Low). 
We constrained relevance to High and Low because our iterative design study revealed that more color--saturation pairs (i.e., 3 hues for Power x 3 saturations for Relevance) were difficult to interpret (Appendix ~\ref{appendix_iterativedesign}).

\subsection{Power Meter: Visualization of the Distribution of Power and Relevance}

\begin{figure}[!h]
    \centering
    \includegraphics[width=0.47\textwidth]{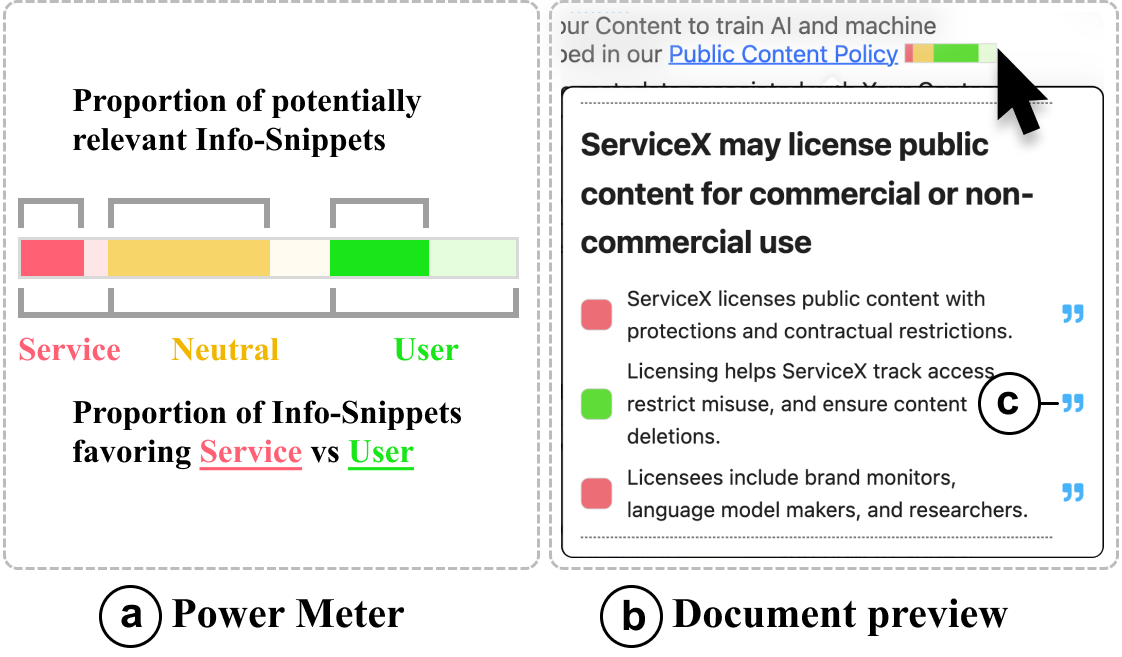}
    \caption{Power Meter visualizes the distribution of power and relevance of Information Snippets within a policy (a). On hover, a preview of Summary Snippets is shown (b) with options to view the referenced Information Snippets (c).}
    \label{fig:powermeter}
    \Description{The figure visualizes the Power Meter. (a) The Power Meter is a horizontal bar that visualizes the distribution of power and relevance of Information Snippets within a policy. (b) When a user hovers over the Power Meter, the system displays the list of Summary Snippets for that policy. (c) Each Summary Snippet is paired with a blue backquote icon that links to the referenced Information Snippet.}
\end{figure}

In TermSight, the policies that are part of the ToS are placed in the top navigation panel. 
To help users form mental models of each policy and decide which policies to read (DG1), each policy in the top navigation panel or hyperlinked within the document is accompanied by a \textbf{Power Meter}: a horizontal bar visualization representing the distribution of Information Snippets in a document, with colors denoting power and relevance (Figure \ref{fig:powermeter}).
Our iterative design study further revealed the need to contextualize Power Meter with concrete previews (Appendix ~\ref{appendix_iterativedesign}). Consequently, hovering over a Power Meter reveals a policy preview containing a list of Summary Snippets (Figure \ref{fig:powermeter}b). Each Summary Snippet in the popup includes a clickable icon (Figure \ref{fig:powermeter}c) that allows users to view the referenced Information Snippet (DG4). 

\subsection{Summary Snippets: Plain language Summaries of Information Snippets}
To help users surface relevant information within a document (DG2) and interrogate the original text (DG4), TermSight provides Summary Snippets accompanied by Highlight Bars for seamless navigation between the summary and the original text (Figure ~\ref{fig:termsight}).

\subsubsection{Summary Snippets}
\label{sec:summarysnippets}
In contrast with document-level ~\cite{tomuro2016automatic, keymanesh2020toward} or section-level summaries~\cite{manor2019plain, august2023paper}, TermSight explores supporting interaction and sensemaking for finer-grained snippets of information within dense sections.
TermSight features \textbf{Summary Snippets}: one-sentence plain language summaries of Information Snippets (\S\ref{sec:infoSnippets}). 
Summary Snippets are designed to be short (detailed in Section \ref{sec:infosummarySnippets}), which have been shown to be more effective for navigational tasks compared to longer snippets ~\cite{cutrell2007you, sweeney2006effective}.
Each Summary Snippet is rendered with color based on the Power and Relevance of the corresponding Information Snippet to provide information  scent~\cite{pirolli1999information} for deciding which snippets to pay attention to (DG2). To enable users to interrogate original text and verify AI-generated summaries (DG4), clicking on a Summary Snippet jumps a user to the corresponding Information Snippet in the document.

\subsubsection{Highlight Bar}
To help users surface and identify Information Snippets while reading or skimming the original text (DG2), TermSight introduces \textbf{Highlight Bars}. Highlight bars are positioned to the left of each Information Snippet in the original text, visually segmenting dense sections into Information Snippets. For paragraphs containing multiple Information Snippets, additional line breaks are added after each snippet for visual separation. Similar to the Summary Snippets, Highlight Bars are color-coded based on the Power and Relevance of the corresponding Information Snippet (\S\ref{sec:powerrelevance}). Users can click on a Highlight Bar to jump to the corresponding Summary Snippet.

\begin{figure*}[!h]
    \centering
    \includegraphics[width=0.97\textwidth]{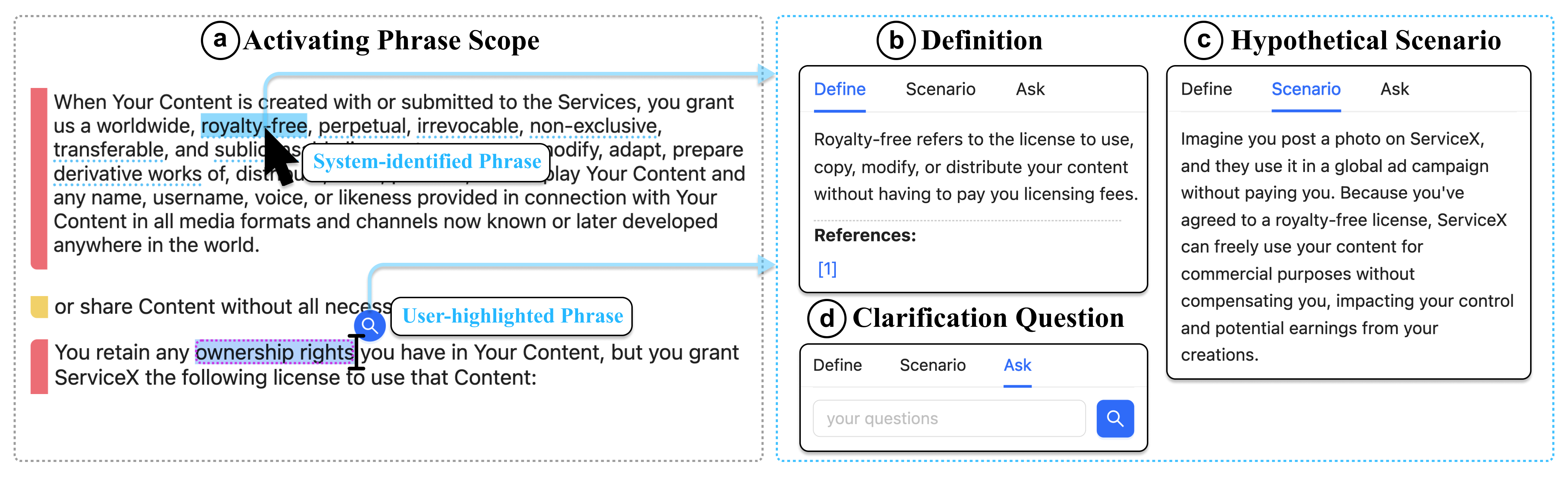}
    \caption{Phrase Scope first identifies jargon or vague phrases in the contract (a). Clicking an identified phrase opens a tooltip with definition (b) and hypothetical scenario (c) while allowing users to ask clarification questions (d).}
    \label{fig:phrasescope}
    \Description{Phrase Scope, a tooltip that provides a term definition and a scenario to contextualize the meaning of a phrase, while also allowing users to ask clarification questions. Users can access Phrase Scope by clicking on a system-identified phrase or highlighting any phrase they do not understand. The figure shows an example output for the phrase `royalty-free’. Definition: `Royalty-free refers to the license to use, copy, modify, or distribute your content without requiring licensing fees.’ Scenario: `Imagine you post a photo on ServiceX, and they use it in a global ad campaign without paying you. Because you agreed to a royalty-free license, ServiceX can freely use your content for commercial purposes without compensating you, reducing your control and potential earnings from your creation.’}
\end{figure*}

\subsection{Phrase Scope: Phrase Identification, Definition, Scenario, and Clarification}

To help users engage with the original text (DG3), \textbf{Phrase Scope} first identifies phrases that may be unfamiliar or ambiguous to general audience readers and underlines them in blue (Figure \ref{fig:phrasescope}). This provides a starting point for users to explore potentially unfamiliar phrases without relying on users to identify them.
Users can also highlight any arbitrary span of text to request Phrase Scope. 
Clicking on the system or user-identified phrase opens Phrase Scope, a tooltip with three components. 
First, the definition of the phrase in the context of the ToS is offered, similar to prior work in other domains~\cite{fok2024qlarify}. 
Beyond definitions, participants in the formative study described the challenge to contextualize abstract phrases in unforeseen scenarios. 
Inspired by prior work that encouraged scientists to reflect on the unintended consequences of their research ~\cite{wang2024farsight, pang2024blip}, Phrase Scope explores the opportunity to prompt reflection on the implications of a phrase by offering hypothetical scenarios personalized to the user persona. Lastly, users have an option to ask clarification questions. More examples of the generated definitions and scenarios can be found in Appendix Table ~\ref{table:example-def-scenario}.

\section{Implementation Details}

TermSight is rendered as a web interface implemented with Next.js. Below, we explain the input, output, processing, and overall performance of TermSight. Additional implementation details and prompts are specified in Appendix ~\ref{Appendix:implementationdetails}.

\subsection{System Input and Pre-processing}
\label{sec:systeminput}

TermSight assumes a set of HTML or markdown source files representing the policies included in the Terms of Service (ToS). We assumed clean source files because the focus of TermSight is not on developing web scraping technologies but rather on investigating meaningful navigational and reading affordances for ToS contracts. Each document is segmented into text \textbf{"chunks"}, composed of one or more paragraphs, which are further vectorized and stored in a vector database (detailed in Appendix ~\ref{Appendix:docpreprocessing}). For features that depend on user preferences, such as classifying the relevance of Information Snippets (\S\ref{sec:infosummarySnippets}) and generating personalized scenarios (\S\ref{sec:PhraseScopeImpl}), TermSight uses a text-based persona that includes users' intended usage of the service (e.g., content consumers vs. content creators) and their values or concerns (e.g., privacy, copyright, etc). The user personas used for the study are further explained in \S\ref{sec:userstudy_materials} and displayed in Appendix ~\ref{Appendix_persona}.

\subsection{\textbf{Obtaining Summary Snippets and Classifying Information Snippets}}
\label{sec:infosummarySnippets}

\begin{figure}[!ht]
    \centering
    \includegraphics[width=0.41\textwidth]{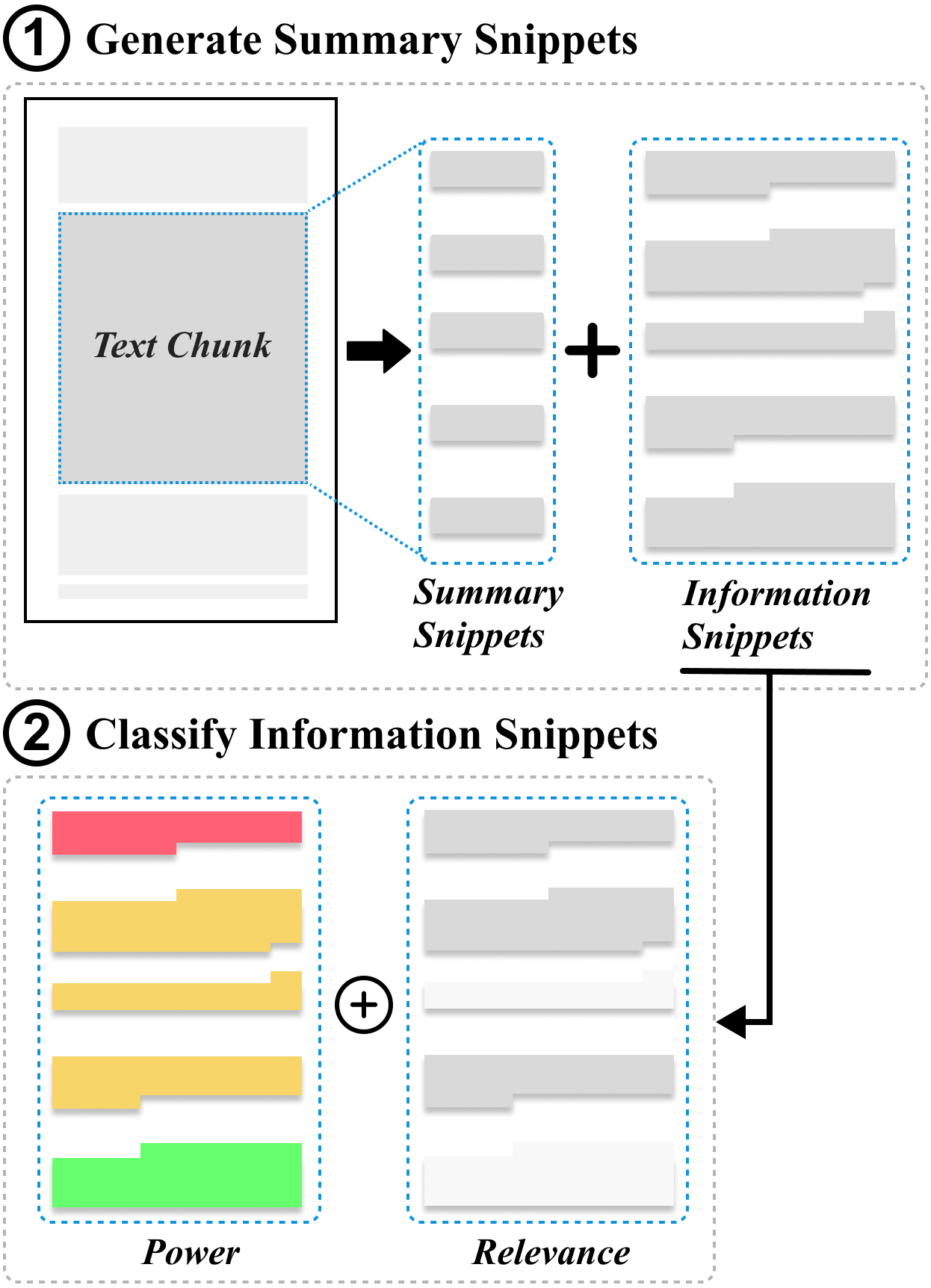}
    \caption{A flowchart of the implementation for (1) obtaining a list of Summary Snippets each referenced to a span of the input text (i.e., Information Snippet) and (2) classifying Power and Relevance for each Information Snippet.}
    \label{fig:summarysnippets}
    \Description{The flowchart of TermSight’s implementation process. (1) The figure shows that a chunk of text is first summarized into a list of Summary Snippets, each referenced to a span of the input text chunk (Information Snippet). (2) After this step, TermSight classifies every Information Snippet with respect to Power and Relevance.}
\end{figure}

For each chunk of text obtained from the document pre-processing pipeline, we prompted GPT-4o to generate a list of one-sentence summaries (Summary Snippets) each referenced to a span of the input text chunk (Information Snippet) as illustrated in Figure ~\ref{fig:summarysnippets}. While the length of the referenced text for each summary is unrestricted, each summary is constrained to a maximum of 12 words based on prompt engineering and prior work showing that short summaries can be more effective for navigation (e.g., 10-20 words) ~\cite{cutrell2007you, sweeney2006effective}. 
We noticed that setting summaries shorter than 12 words led to excessive fragmentation (i.e., too many Summary Snippets), while longer summaries tried to cover too much information, making them less skimmable. 
We then prompted GPT-4o to classify each Information Snippet along two dimensions: Power and Relevance (to user persona). Prompts can be found in Appendix ~\ref{Appendix:obtainsummary} and ~\ref{Appendix:classifyinfos}. 

\subsection{\textbf{Generating Phrase Scope}}
\label{sec:PhraseScopeImpl}

Similar to prior work \citep{guo2023personalized, fok2024qlarify}, we prompt an LM to identify potentially unfamiliar and ambiguous phrases within document chunks produced by the pre-processing pipeline.
To generate in-context definitions for the identified phrases, we used a retrieval-augmented question answering approach ~\cite{fok2024qlarify}. We vectorize the document chunks (\S\ref{sec:systeminput}) and query these chunks to retrieve relevant document context. 
These chunks are collectively referred to as \textbf{"retrieved chunks"}. Then, we prompted GPT-4o to generate an in-context definition with references to the retrieved chunks. When users ask additional questions, the same retrieval-augmented question answering pipeline used for generating definitions is applied, with the only difference being the question asked. 
Scenarios aim to prompt reflection on the potential implications of a phrase based on users' intended usage of the service and personal value. 
We employed GPT-4o with zero-shot prompting to generate the scenarios tailored to the user persona. 
The prompts used for Phrase Scope are detailed in Appendix ~\ref{Appendix:phrasescope}.

\subsection{Evaluation of System Output}

Before reporting on our user study, we randomly sampled and conducted a manual evaluation of TermSight's core outputs (e.g., summary snippets, definitions, scenarios). Our goal was not to assess advances in system performance, but to verify that the system can produce meaningful outputs that could support ToS reading. Across 116 sampled Information Snippets, we found 5 imperfect classifications of power or relevance due to the lack of full context in the input Information Snippet. Similarly, we reviewed 113 generated definitions and scenarios each. One scenario was found to be factually incorrect due to a hallucination. 
The input context and phrase stated that users do not gain ownership rights by downloading content from the service. However, the scenario claimed that users might lose ownership rights over the content they create by uploading it to the service.
All the definitions were correct except for 4 overly general definitions for service-specific phrases not explicitly defined anywhere in the ToS. More details of the evaluation are provided in Appendix~\ref{appendix_system_eval}.

\section{User Study}
Similar to prior HCI studies ~\cite{chen2023marvista, cao2023dataparticles, kambhamettu2024explainable}, the goal of our user study is not to prove TermSight as the definitive solution through a series of ablation experiments, but to use TermSight to explore the problem and solution space of designing contract reading interfaces.
We asked the following research questions:

\begin{quote}

\textbf{RQ1}: How did participants perceive TermSight and its features?

\textbf{RQ2}: How did TermSight influence ToS comprehension and recall?

\textbf{RQ3}: How did participants read with TermSight and its features?

\end{quote}

\subsection{Study Design}
We conducted a within-subject study where each participant used both TermSight and a baseline interface to read two services' ToS, once with each interface variant. The order of the interface and service type was counterbalanced to reduce ordering effects. 

\subsubsection{Baseline}

Because there is no well-established standard ToS reader, we use the most commonly used interface as the baseline (i.e., an HTML reader).
The baseline interface had the same layout as TermSight without the Power Meter, Summary Snippet, Highlight Bar, and Phrase Scope (Appendix Figure ~\ref{fig:termsightbaseline}). In place of the Summary Snippets was a table of contents that can be used to navigate to different sections.
Participants' feedback in the user study confirmed that the baseline offered a strong frame of reference (\S\ref{sub:RQ1intro}).
While reading interfaces exist for other domains (e.g., ~\cite{august2023paper, chen2023marvista, fok2022scim}), adopting them to contracts would result in novel systems unsuitable as baselines. For example, prior work has designed key questions~\cite{august2023paper} or faceted highlighting~\cite{fok2022scim} specific to academic papers, which require co-design studies with lawyers to adapt them to contracts. 
Injecting additional features (e.g., overview summaries ~\cite{tabassum2018increasing}, chat~\cite{zhao2020talk}) into one or both conditions may also introduce confounds unrelated to our research questions. 
For example, by adding a policy or section-level summary in the baseline, the comparison with TermSight will be confounded by the design of summaries in the baseline (e.g., levels of complexity and detail), a variable that may impact different readers differently \cite{august2024know}.

\subsubsection{Materials}
\label{sec:userstudy_materials}

\paragraph{Terms of Service}
For the user study, we used the ToS of one social media site (Reddit) and one e-commerce site (Poshmark) because social media and e-commerce platforms are among the most visited digital services and are representative of the standard long-form contract used in prior studies ~\cite{kay2010textured, taber2020beyond, good2005stopping, tabassum2018increasing, obar2020biggest}. For each service, we collected policies that are linked as part of the ToS. We did not include location-specific policies (e.g., a California Privacy Notice). For Reddit, we also did not include policies for advertisers, publishers, or governmental agencies to keep the number of policies for both services similar. We collected a total of 14 policies for Reddit and 15 for Poshmark. The service names for Reddit and Poshmark were anonymized as ServiceX and ServiceY. 

\paragraph{User Persona}
Features of TermSight rely on a user persona to determine relevance. For the user study, we designed two personas based on users' information needs identified in the formative study: a content consumer who posts personal content on social media sites and a buyer who rarely posts reviews on e-commerce sites (Appendix ~\ref{Appendix_persona}).
Participants in the user study validated and found their given user persona for both services to highly align with their personal usage of the service and their personal value (Figure ~\ref{fig:userstudyparticipant}).


\subsubsection{Study Procedure}

The entire study took 90 minutes, and participants were compensated 25\$. The study was approved by the Institutional Review Board (IRB) at our organization.
The study was composed of two reading sessions, 45 minutes each. After obtaining consent and before the reading sessions, we asked participants about their past experiences with ToS. Within each reading session, participants first completed a pre-survey and reported their past experiences interacting with social media or e-commerce platforms. Then, an anonymized description of the service and user persona was given. Similar to prior studies~\cite{august2023paper}, participants were given 10 minutes to read the ToS without having to speak aloud. Reminders were given at the 5 and 1-minute mark. After the reading session, participants completed seven 5-point Likert-style rating questions about their reading experiences, followed by one free-response recall question and six multiple-choice comprehension questions. We also conducted semi-structured interviews with participants about their experiences and reading strategies for 10 minutes. The same procedure was repeated for the second reading session. The study materials are included in Appendix ~\ref{Appendix_comparativestudy}.

\subsection{Participants and Setting}
We recruited 22 participants through Prolific. The recruitment required participants to be over 18, fluent in English, located in the US, not have color deficiencies, and not be legal professionals. We did not require participants to have prior experience in reading ToS. Two participants encountered technical difficulties, and their data were removed from the analysis. 
The median of participants’ self-rated familiarity with ToS was 3.5 (\textit{\(\sigma=1.5\)}) out of 5. 
Most participants were generally familiar with social media and e-commerce platforms and found their given user persona to highly align with their own profile (Figure ~\ref{fig:userstudyparticipant}). None of the participants had read or remembered the ToS for Reddit and Poshmark prior to the study. Additional participant demographics can be found in Appendix ~\ref{Appendix_demographics}.

\subsection{Measures}

\subsubsection{Reading Experiences}

We collected seven 5-point Likert-style ratings (1=“Not at all,” 5=“Very”) of participants' reading experiences (Appendix ~\ref{Appendix_likert}). The experience measures included ease of reading (adopted from perceived effort measure in NASA TLX) ~\cite{august2023paper, hart1988development}, perceived understanding ~\cite{august2023paper}, and confidence in obtaining relevant information ~\cite{august2023paper}. 
We added two questions specific to challenges identified in our formative study: the ease of deciding which policy to read and what text to read within a policy. Finally, we included two questions on participants' willingness to read ToS with the interface and their willingness to spend more time on the ToS with the interface~\cite{august2024know}.

\subsubsection{Comprehension}

We wrote 10 multiple-choice questions for each service and selected 6 questions that are relevant for the given user persona (\S\ref{sec:userstudy_materials}) for each service, similar to prior work on evaluating reading comprehension for ToS and privacy policies~\cite{tabassum2018increasing, vu2007users, august2023paper}. 
We designed the comprehension questions to have one answer based on the original text, independent of features of TermSight (e.g., Summary Snippets). Consequently, participants need to read the original text to fully answer the questions. 
The questions were different for the two services to minimize learning effects. Comprehension was measured as the number of questions participants got right. 
We note that past studies with reading interfaces often fail to show differences in how people answer comprehension questions~\cite{tabassum2018increasing, korunovska2020challenges, robinson2020beyond, august2023paper, head2021augmenting}. Similar to prior work ~\cite{august2023paper}, our intention in including comprehension questions was to ensure that TermSight did not detract from overall comprehension. The questions can be found in the supplemental materials.

\subsubsection{Recall}

For recall, participants were asked to write down information in ToS they found interesting, surprising, or any detail they remembered: \textit{``What did you learn from reading the ToS? Recall one or more interesting things you learned from the ToS.''} To analyze these free-form responses, we manually broke the response into single references to a clause in the ToS. Then, we labeled whether each reference was correct based on the ToS. Recall is measured as the number of correct facts in the response ~\cite{taber2020beyond, good2005stopping}.

\subsubsection{Feature Usage}
To understand users' interaction behavior, we logged all viewport scrolls and feature usage during the reading session along with timestamps. This allows us to measure the frequency of feature usage of TermSight and the baseline interface.

\subsection{Analysis}

In this section, we introduce our analysis framework for the user study. We used a causal framework to understand the effects of the treatment and a Bayesian analysis to estimate the treatment effect (\S\ref{sub:Bayesian Analysis}). Then, we discuss the thematic analysis (\S\ref{sub:Thematic analysis}) used to analyze the qualitative responses from the semi-structured interviews.

\begin{figure}[!ht]
    \centering
    \includegraphics[width=0.37\textwidth]{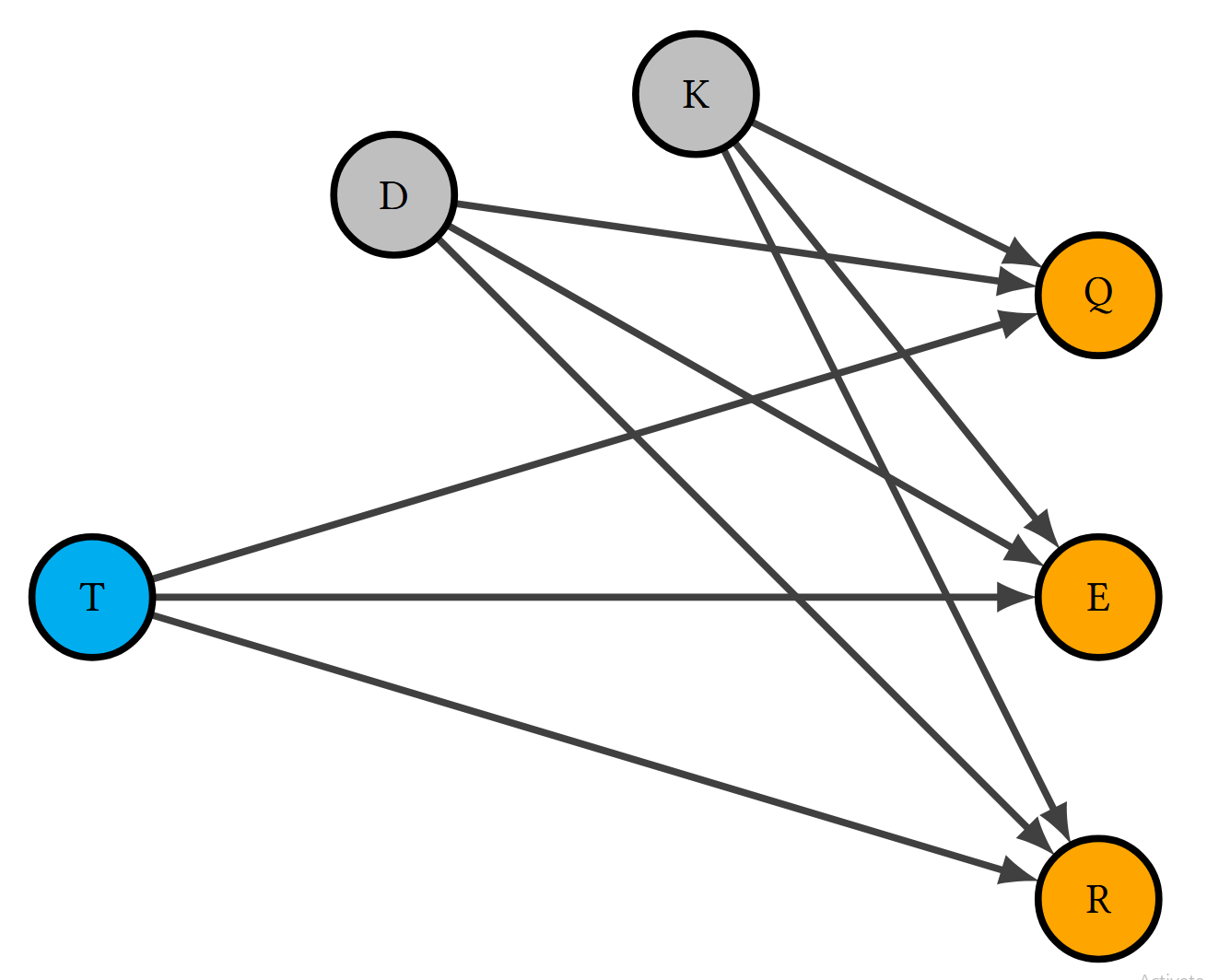}
    \caption{The causal graph (DAG) has three types of nodes: treatment ($T$); pre-treatment covariates $D$ (demographics), $K$ (knowledge of Terms of Service); and measured outcomes $Q$ (comprehension), $R$ (recall), and $E$ (user experience). Since we use randomized assignment for the interface type (treatment vs. control), ToS type (social media vs. e-commerce), and presentation order (first vs. second), the treatment $T$ is independent of the pre-treatment covariates $D$ and $K$. To estimate the causal effect of the treatment $T$ on the outcomes $Q$, $E$, and $R$, we can use a simple regression model, without conditioning on covariates $D$ and $K$, since there is no backdoor path from the treatment $T$ to the outcome variables.}
    \label{fig:dags}
    \Description{The causal graph (DAG) has three types of nodes: treatment (T) colored blue; pre-treatment covariates D (demographics), K (knowledge of Terms of Service) colored gray; and measured outcomes Q (comprehension), R (recall), and E (user experience) colored orange. Since we use randomized assignment for the interface type (treatment vs. control), ToS type (social media vs. e-commerce), and presentation order (first vs. second), the treatment T is independent of the pre-treatment covariates D and K. To estimate the causal effect of the treatment T on the outcomes Q, E, and R, we can use a simple regression model, without conditioning on covariates D and K, since there is no backdoor path from the treatment T to the outcome variables.}
\end{figure}

\subsubsection{Bayesian Analysis}
\label{sub:Bayesian Analysis}
First, we used a structural causal framework popularized by~\citet{Pearl2009} to understand the effects of the treatment (Figure ~\ref{fig:dags} shows the DAG for the study). More details about structural causal framework can be found in Appendix ~\ref{sub:Structural Causal Framework}.
Next, we used Bayesian inference to estimate the treatment effect of TermSight on the outcomes of interest. While the use of Bayesian estimates is growing in HCI~\cite{Kay2016,Koshy2023}, we briefly justify its use over traditional non-Bayesian methods. 
First, as \citet{Kay2016} points out, a Bayesian framework leads to an accumulation of knowledge within HCI, where the posterior of the parameters in a prior work can serve as the prior in the current experiment. Second, a Bayesian model is transparent---the researcher will foreground all the assumptions in the analysis via their model. Third, use of a Bayesian framework shifts the discussion from ``did it work'' to ``effect size of the intervention''~\cite{Kay2016}. While NHST techniques can compute the confidence interval, these intervals are susceptible to misinterpretation~\citep{Belia2005,Hoekstra2014} and, importantly, underemphasized in favor of $p$-values. Finally, as ~\citet{McElreath2015} points out, a Bayesian model with the use of maximum entropy priors for the parameters (e.g., the normal distribution) is \textit{the most conservative} given the evidence to estimate the effect of the treatment.

\subsubsection{Thematic Analysis}
\label{sub:Thematic analysis}

To analyze the qualitative responses from the semi-structured interviews, we adopted the six phases of reflexive thematic analysis suggested by Braun and Clarke ~\cite{braun2006using}, following the same procedure as our formative study (\S\ref{sec:needfinding_analysis}).

\section{Findings}
\label{sec:Findings}

\subsection{RQ1: How did participants perceive TermSight and its features?}
\label{sub:RQ1intro}

17 participants highlighted that the baseline interface was easier to read than most of the ToS they had seen in the past, which often lacked the top navigation panel or a table of contents.
When asked about which version of the interface they prefer, 19 participants preferred TermSight. One participant preferred the baseline interface, citing difficulty navigating TermSight’s two scrollable columns on a small screen with a trackpad that lacks scroll functionality. 
Below we present our quantitative findings on the experience outcomes (\S\ref{sub:Quantitative Findings, Experience}), followed by qualitative findings organized around recurring themes about features of TermSight (\S\ref{sub:QualFindings-powermeter} -- \S\ref{subsubsec:phrasescopevalue}).

\subsubsection{TermSight improved all seven user experience measures}
\label{sub:Quantitative Findings, Experience}

\begin{figure*}[!ht]
    \centering
    \includegraphics[width=0.85\textwidth]{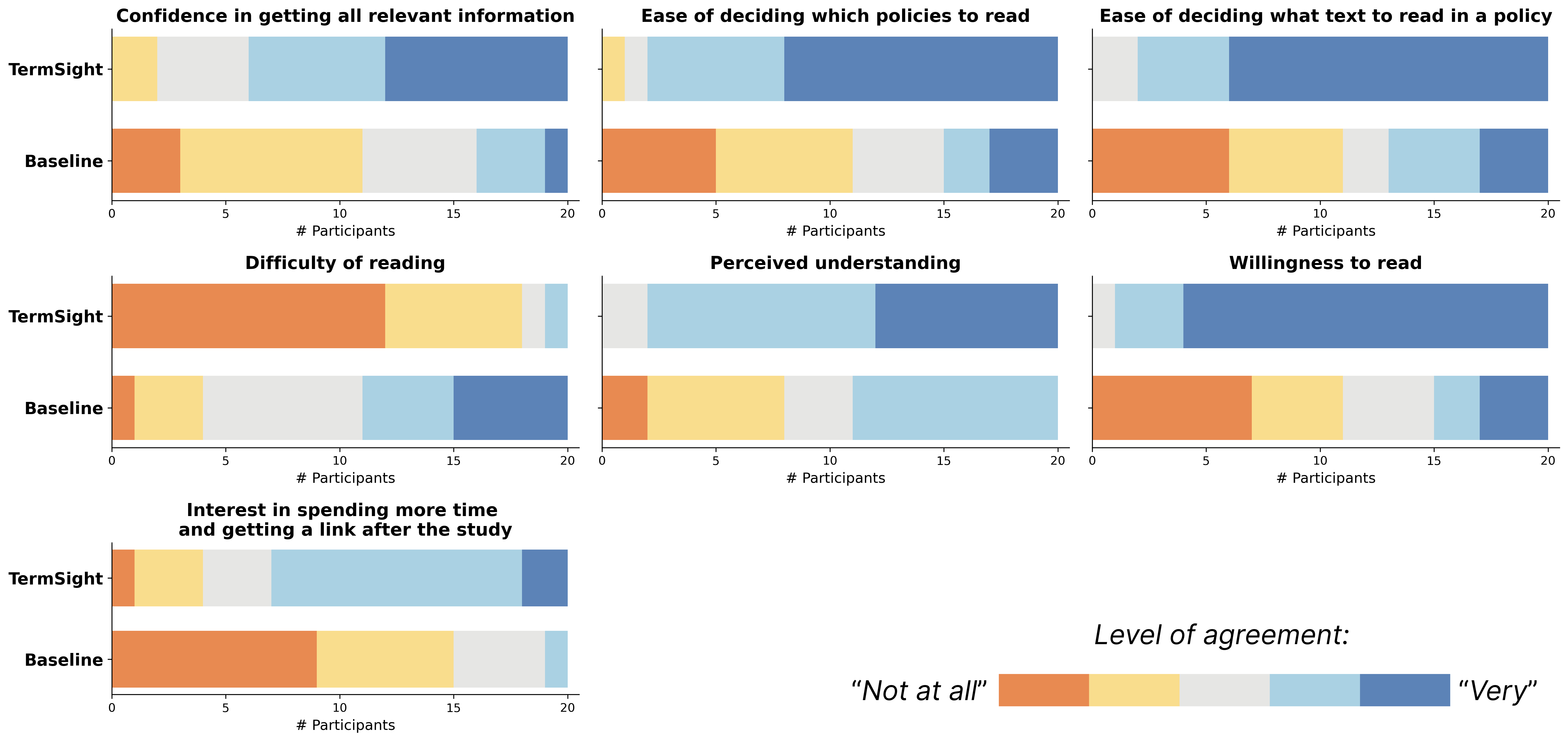}
    \caption{Participants' ratings of reading experiences. Participants found it to be easier to find and understand relevant information with TermSight. Moreover, participants were more willing to read and spend more time on ToS with TermSight.}
    \label{fig:experiencelikert}
    \Description{The figure shows seven barplots of 5-point Likert Ratings of participants' reading experiences with TermSight and the baseline interface. Participants rated TermSight to be better for all seven experience measures. For example, participants found it to be easier to find and understand relevant information. Participants were also more willing to spend more time on the ToS with TermSight and read other ToS with TermSight.}
\end{figure*}

Figure ~\ref{fig:experiencelikert} shows participants' ratings of the 7 reading experience measures for both interfaces. 
Our Bayesian analysis (model details in Appendix~\ref{sub:Modeling Experience Outcomes}) reveals a significant effect of the TermSight interface on the reading experience measures. 
Participants found reading ToS with TermSight to require significantly less effort. When navigating ToS, deciding which policy to read and what text to read within a policy was significantly easier with TermSight. Moreover, participants were significantly more willing to read ToS and spend more time reading ToS with TermSight.
The forest plots shown in Appendix Figure ~\ref{fig:forest_comprehension} demonstrate that the 94\% Highest Posterior Density Interval (HPDI\footnote{In Bayesian analysis, the 94\% HPDI refers to that smallest interval of the posterior distribution that has 94\% of the probability mass. It is common in Bayesian analysis to use intervals distinct (97\% and 89\% HPDI intervals are also used) from the typical 95\% value to avoid the confusion of the frequentist confidence intervals.}) for each treatment--control pair for every user experience measure does not overlap with each other. This indicates a significant positive effect of the treatment on all 7 user experience measures. 

\begin{figure}[htbp]
    \centering
    \includegraphics[width=0.47\textwidth]{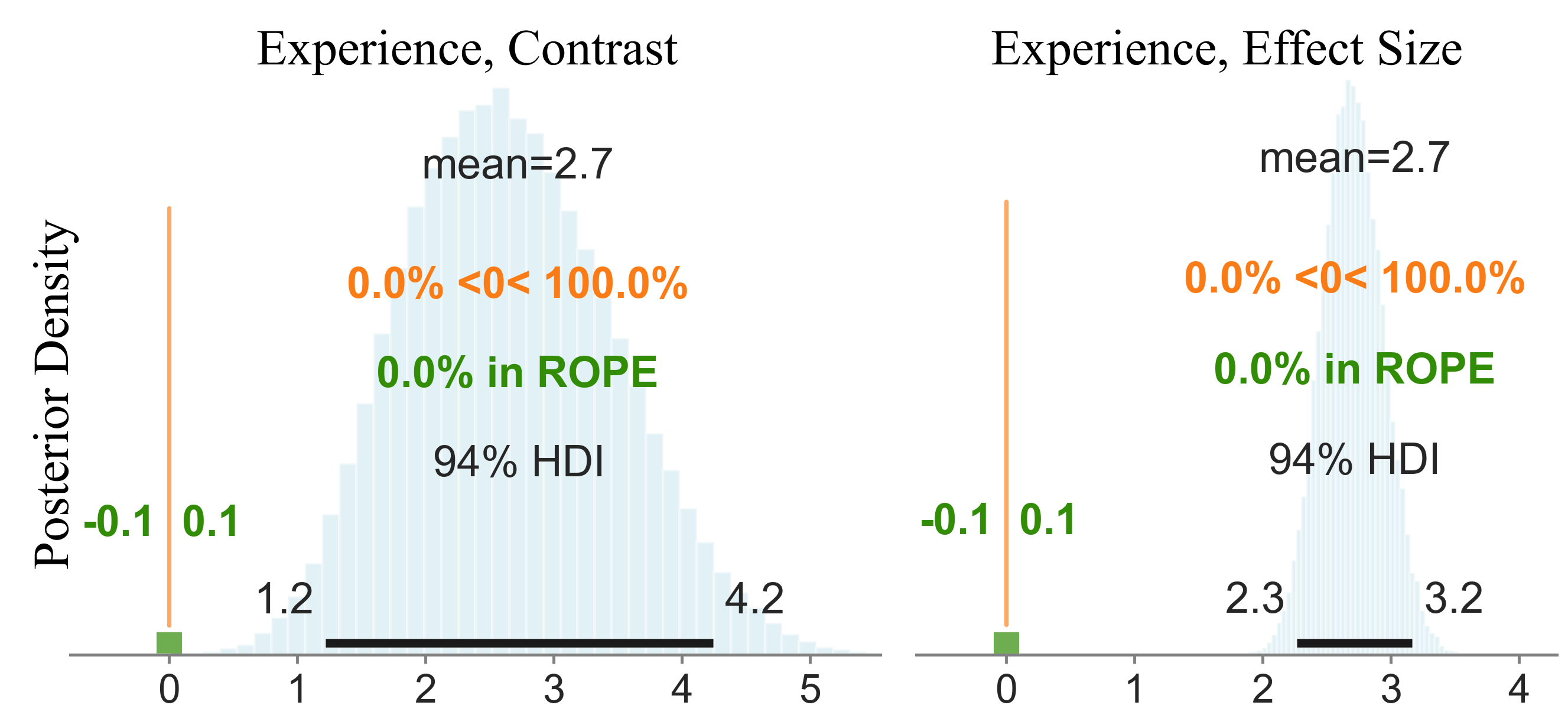}
    \caption{The posterior distribution of the contrast between treatment and control on user experience is shifted to the right, with the posterior distribution of the effect size centered around 2.7 without overlapping with ROPE, suggesting a significant and large effect size.}
    \label{fig:posterior_experience}
    \Description{Figure 7 shows the posterior distribution of the contrast between treatment and control on user experience. Note that the posterior distribution of the contrast is shifted to the right, with the posterior distribution of the effect size centered around 2.7 without overlapping with ROPE, suggesting a significant and large effect size.}
\end{figure}

In Figure ~\ref{fig:posterior_experience}, we show the contrast between the treatment and control conditions averaged across 7 user experience measures as well as the posterior of the effect size. 
The posterior distribution of the effect size (Cohen's $d$) of the treatment on user experience is centered around 2.7, indicating a large effect size\footnote{This is the effect size measured not on the outcome Likert scale (1--5) but on the latent scale of the model coefficients.}. Because the HPDI doesn't overlap with the ROPE\footnote{We use a Region of Practical Equivalence (ROPE) of [-0.1, 0.1]. ROPE is a region where the difference between the treatment and control conditions is considered practically equivalent. This prevents us from focusing on small differences that are not practically meaningful.}, the effect of TermSight on user experience is significant.
To further disentangle how individual features of TermSight contributed to the improvement in user experience, we present recurring themes related to participants' perceptions and feedback on features of TermSight.

\subsubsection{Power Meter reduced navigation cost}
\label{sub:QualFindings-powermeter}

All 20 participants described that the Power Meter helped them gain an intuition of the content within each policy and supported decision-making on which policy to pay attention to.

\begin{quote}
    \textit{``The color system was really guiding a lot of my decisions on which policies to read. The vibrant red was always what I try to get to first and the things that I cared the most about. So, finding which terms I'm agreeing to that I have the least amount of agency.''} (P12)
\end{quote}

P6 used the TermSight system first. However, he lamented the fact that when using the baseline interface without the Power Meter, deciding which policy to read becomes a guessing game.

\begin{quote}
    \textit{``Policies like Terms of Service or privacy policy are kind of broad. Does that apply to something I would actually care about? I don't know. So I'd have to click each one and have to go through it. It was more like a guessing game.''} (P6)
\end{quote}

Additionally, eight participants described how the document preview reduced navigation cost by allowing them to access information from other policies without \textit{``clicking on the link and possibly losing where I am on''} (P1).

\subsubsection{Summary Snippets and Color provided guidance and helped surface power and relevance within documents}

Eighteen participants described how the color of the Summary Snippets helped surface where power lies within a document, which is challenging to ``see'' otherwise. 

\begin{quote}
    \textit{``The color coding gave me a better understanding of who's benefiting more, how it's benefiting them, and how it affects me. Versus [in the baseline], it feels like it only really benefits the company itself.''} (P18)
\end{quote}

Moreover, all participants described how the color and summary allowed them to make decisions and prioritize what to focus on strategically.

\begin{quote}
    \textit{``I really liked having the summary with colors. I had a hierarchy of how I was going to read stuff. I made sure I read all the [saturated] red ones, and then the  [saturated] green. If I had time, I would skim the [saturated] yellow and see if anything stuck out. It would take me a lot longer to filter for information otherwise.''} (P6)
\end{quote}

While participants described focusing on the more saturated colors, especially red and green, six participants complimented how TermSight made AI decisions (i.e., classification of power and relevance) transparent by keeping all the Summary Snippets visible and described skimming the less saturated ones to validate and make sure they didn't miss information.

\subsubsection{Summary Snippets simplified and broke down dense text to support sensemaking}

Summary Snippets \textit{``broke down the entire section into a few bullet-pointed summaries''} (P11), which helped participants surface and absorb relevant information hidden in dense text (20/20). Nine participants further described how the Summary Snippets simplified the text and served as a scaffold to encourage reading the original text. 

\begin{quote}
    \textit{``Not only does it [TermSight] bring me up to things that I would normally miss. but then that would be an introduction or a guide, or a push towards reading the whole 3 paragraphs. ... It is really guiding me through this entire document, and it's simplifying it.''} (P2)
\end{quote}

P20 described the experience of using Summary Snippets to make sense of the big ideas of the section while being able to dive deeper as forming \textit{``a web of thought and ideas: what everything kind of means together and piece by piece''}.
Not only do the Summary Snippets serve as an entry point to the original text, four participants explicitly highlighted how the Summary Snippets helped them check their understanding after reading the original text.

\subsubsection{Term definitions and scenarios helped consume original text and envision implications}
\label{subsubsec:phrasescopevalue}

Fourteen participants noted that the system identified phrases attracted their attention and prompted them to check on their understanding of the phrase that they may not have noticed otherwise. Moreover, the definitions and scenarios made legal language more approachable for participants (14/20). 

\begin{quote}
    \textit{``It really just made it a lot easier to understand, because most of the time they write it in legal terms that most people don't know. People don't speak legal language. So this broke it down easier for most people like the Layperson.''} (P17)
\end{quote}

12 participants highlighted how the scenarios helped envision potential implications of the phrase in the context of the user's intended usage of the service and was \textit{``easier to resonate''} (P17). 

\begin{quote}
    \textit{``I like that it provides a real-world example of when and how [the phrase] would be relevant to someone. I think it actually does better explaining the concept than the definition. Because it's not just the definition. It's the definition in context.''} (P11)
\end{quote}

Moreover, one participant (P1) noted how the scenarios helped her come up with more questions, which she asked with the Ask function. Through this process, TermSight helped P1 to quickly obtain information from related policies.

\subsection{RQ2: How did TermSight influence comprehension and recall?}

\begin{figure*}[htbp]
    \centering
    \begin{subfigure}[t]{0.315\textwidth}
        \centering
        \includegraphics[width=\linewidth]{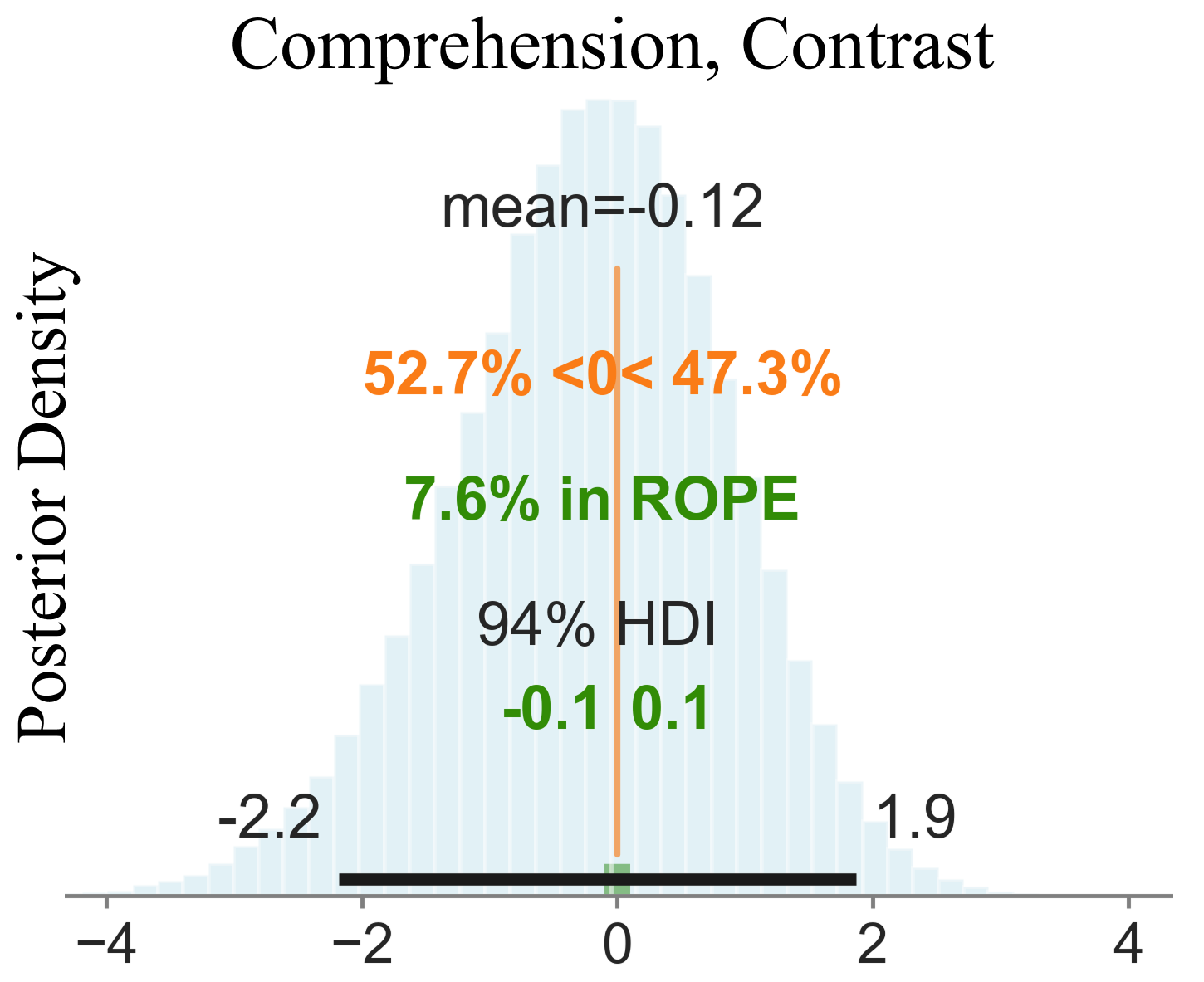}
        \caption{Posterior distribution of contrast (between treatment and control) on comprehension.}
        \label{fig:posterior_comp}
        \Description{The figure shows the posterior distributions of the contrast between the treatment and control conditions on comprehension outcome. The posterior distribution is nearly evenly split around zero and overlaps with ROPE, indicating no effect of the treatment on the comprehension quiz questions.}
    \end{subfigure}\hspace{3em}
    \begin{subfigure}[t]{0.4\textwidth}
        \centering
        \includegraphics[width=\linewidth]{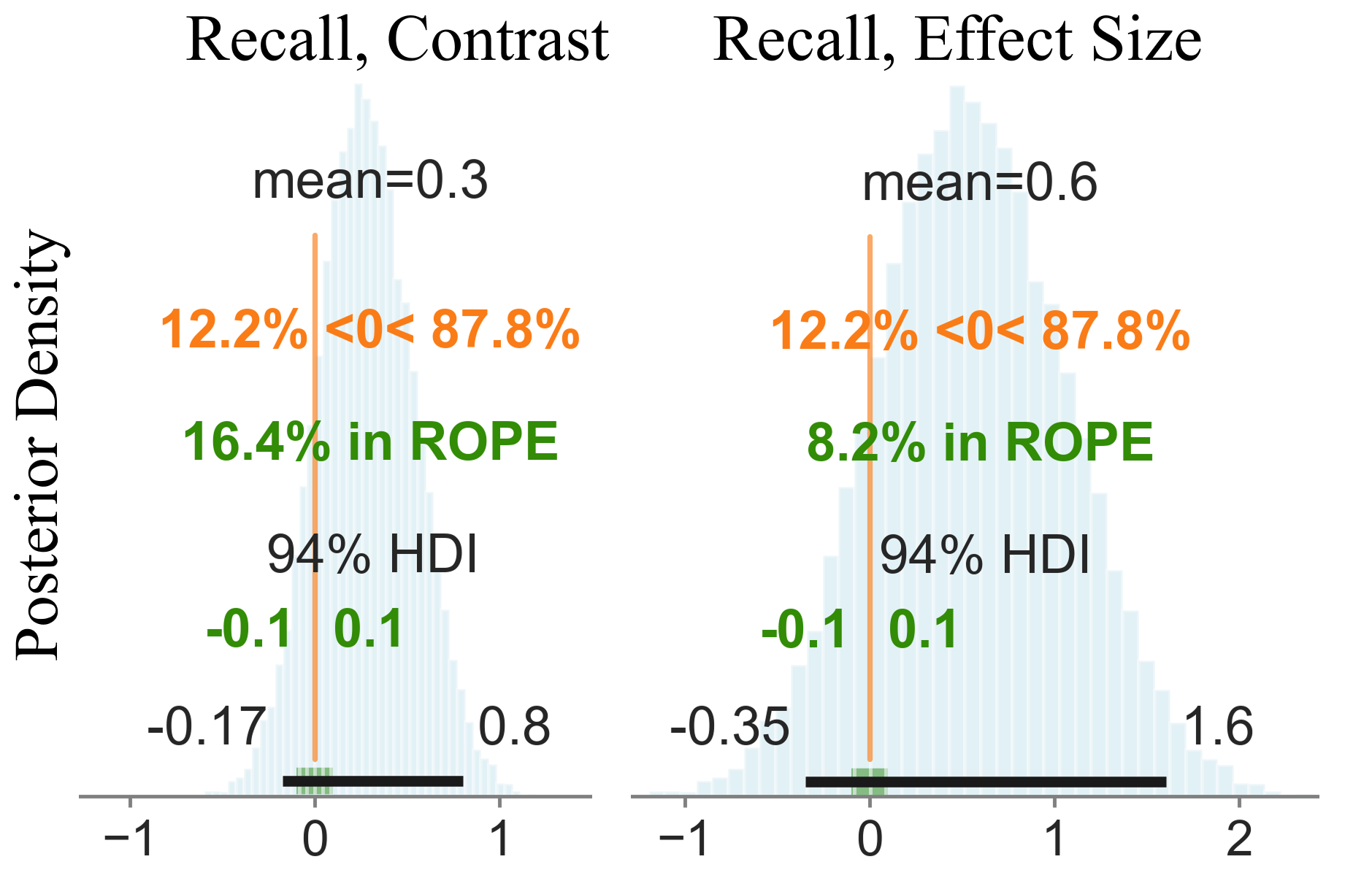}
        \caption{Posterior distribution of contrast (between treatment and control) and effect size on recall.}
        \label{fig:posterior_recall}
        \Description{On the left, the figure shows the posterior distributions of the contrast between the treatment and control conditions for the recall outcomes. The posterior distribution of the treatment condition is shifted to the right, indicating a positive influence of the treatment on Recall. On the right, the figure shows the posterior distributions of the effect size of treatment centered around 0.6, suggesting a medium effect size. Yet, the HPDI has an overlap with the ROPE of [-0.1, +0.1], implying no significant effect of the treatment on the user recall.}
    \end{subfigure}

    \caption{In (a), the distribution centers around zero and overlaps with ROPE, indicating no effect of treatment on comprehension. (b) illustrates a positive influence of treatment on recall with the posterior distribution of the effect size centered around 0.6, suggesting a medium effect size. Because the HPDI overlaps with ROPE, there is no significant effect of treatment on recall.}
    \label{fig:posterior_comp_recall}
    \Description{In (a), the distribution centers around zero and overlaps with ROPE, indicating no effect of treatment on comprehension. (b) illustrates a positive influence of treatment on recall with the posterior distribution of the effect size centered around 0.6, suggesting a medium effect size. Because the HPDI overlaps with ROPE, there is no significant effect of treatment on recall.}
\end{figure*}

\subsubsection{Comprehension}
\label{subsec:comprehensionresults}

Out of 6 questions, participants on average scored 2.00 (\textit{\(\sigma=1.03\)})  when using TermSight and 2.15 (\textit{\(\sigma=1.23\)}) when using the baseline interface. 
Our Bayesian analysis for the comprehension quiz (model details in Appendix ~\ref{sub:Modeling Comprehension Outcomes}) reveals no significant differences in the comprehension scores across interface conditions. 
As shown in Figure ~\ref{fig:posterior_comp}, the posterior distribution of the contrast between treatment and control is centered around zero. 
When the question ID and service type are fixed, there are significant overlaps between the 94\% HPDI for each treatment--control pair as shown by the forest plots in Appendix Figure \ref{fig:forest_quiz}, suggesting no significant differences.
While we aimed to design comprehension questions with one single answer based on the original text, we acknowledge that legal contracts and problems may contain room for interpretation ~\cite{marotta2025building}.
Therefore, we performed a supplemental, more conservative analysis by removing any question where the applicability of the original text to the question may contain room for interpretation, potentially leading to different answer choices (removed 2 out of 6 for each service type).
The findings remained consistent, showing no significant difference in comprehension. More details of the analysis can be found in Appendix \ref{sub:Additional_comp_analysis}.\enlargethispage*{12pt}

\subsubsection{Recall}
\label{subsec:recallresults}

Participants on average recalled 1.54 (\textit{\(\sigma=1.96\)}) correct facts using TermSight and 1.30 (\textit{\(\sigma=1.72\)}) correct facts using the baseline interface. 
Our Bayesian analysis for the recall task (model details in Appendix \ref{sub:Modeling Recall Ouctomes}) reveals no significant differences in the recall scores across interface conditions. 
Figure ~\ref{fig:posterior_recall} shows the posterior distribution of the effect size of the treatment on recall ($\mu=0.6$, HPDI = [-0.35, 1.6]). Despite the mean being 0.6, suggesting a medium effect size, the HPDI interval has an overlap with the ROPE, implying no significant effect of the treatment on recall.

\subsection{RQ3: How did participants read with TermSight and its features?}

\begin{figure}[!ht]
    \centering
    \includegraphics[width=0.47\textwidth]{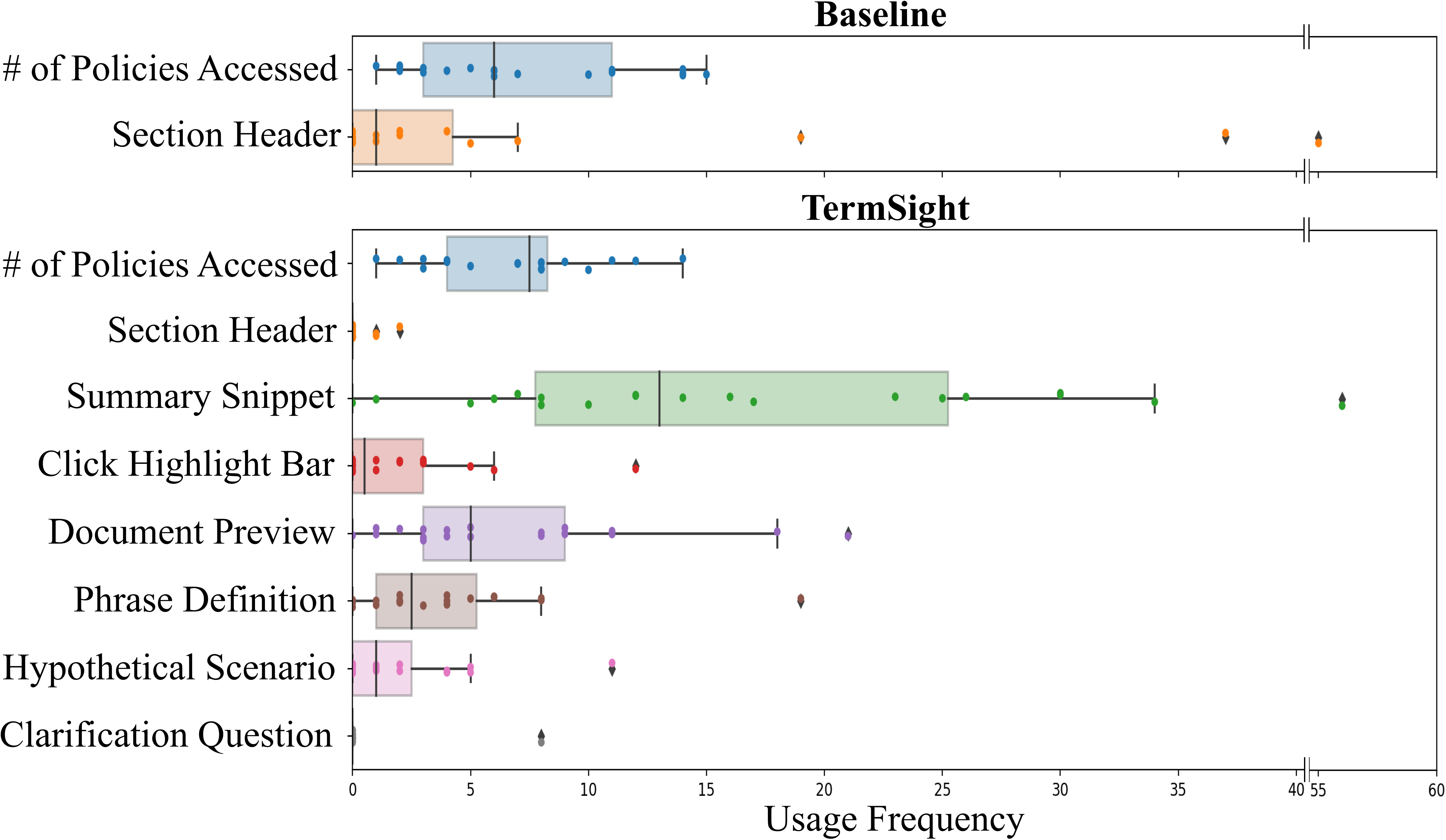}
    \caption{Feature usage. Each dot represents the number of times a feature is used by one participant during a reading session. Participants used most features of TermSight. In contrast to the baseline, participants, when using TermSight, more frequently clicked the Summary Snippets as opposed to the section headers to navigate to the original text.}
    \label{fig:featureusagecount}
    \Description{Box plots for the usage of features during the reading session for TermSight and the Baseline interface. Participants used most features of TermSight. In contrast to the baseline, participants, when using TermSight, more frequently clicked the Summary Snippets as opposed to the section titles to navigate to the original text.}
\end{figure}

\subsubsection{Feature Usage}

In the baseline interface, participants on average navigated to 7 policies (\textit{\(\sigma=5\)}). Participants described trying to skim everything from top to bottom and often skipped difficult sections. 11 participants clicked the section heading in the table of contents at least once to navigate to targeted sections.

\begin{figure}[!ht]
    \centering
    \includegraphics[width=0.4\textwidth]{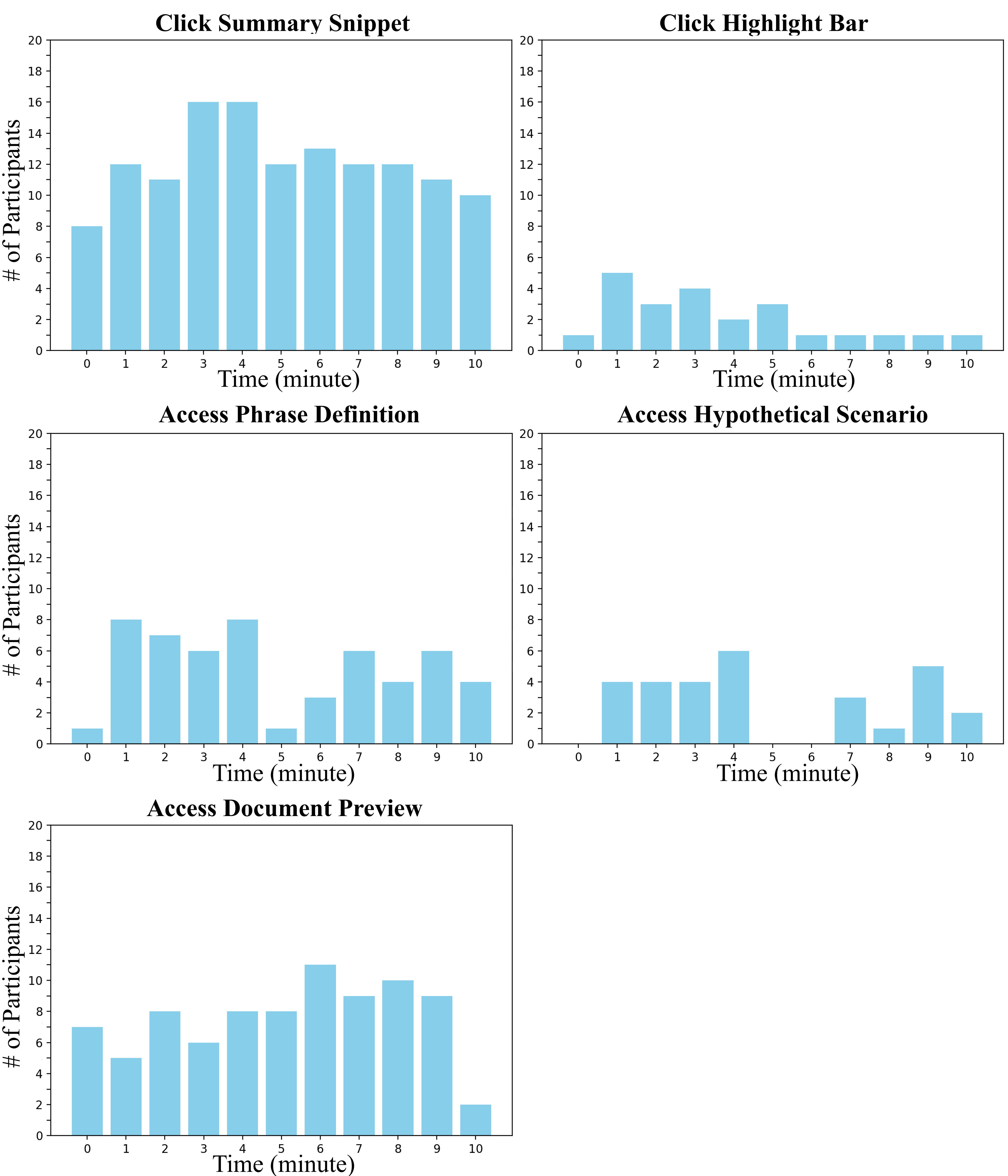}
    \caption{Minute-by-minute usage of features of TermSight during the ten-minute reading task. Participants used these features throughout the reading session, rather than being limited to the beginning or end. The Summary Snippet is the most frequently used feature throughout.}
    \label{fig:featureusagetime}
    \Description{Five barplots of the minute-by-minute usage of features of TermSight during the ten-minute reading task (i.e., Accessed Document Preview, Click Summary Snippet, Click Highlight Bar, Accessed Phrase Definition, and Accessed Phrase Scenario). Participants used these features throughout the reading session, rather than being limited to the beginning or end. The Summary Snippet and Document Preview are the most frequently used features throughout.}
\end{figure}

As shown in Figure ~\ref{fig:featureusagecount} and ~\ref{fig:featureusagetime}, participants used most features of TermSight throughout the session. When using TermSight, participants on average navigated to 6 policies (\textit{\(\sigma=3\)}). 19 participants hovered over a Power Meter to gain an overview of documents in ToS for more than 1 second. 13\% (17/135) of the total usage was for previewing documents hyperlinked inline, demonstrating that the Power Meter provided value also for inline hyperlinks. 

Summary Snippets were the most frequently used feature. 19 participants clicked the Summary Snippets to navigate and read the original text at least once (\textit{\(\mu=17, \sigma=13\)}). In contrast, participants rarely clicked the section headers to navigate to the original text (4 clicks in total across 3 participants). This demonstrates that Summary Snippets were potentially more useful as navigational features to surface and interrogate relevant information from the original text compared to section headers. In addition, 10 participants clicked the Highlight Bar to read the Summary Snippet after reading the original text to check their understanding. 

For Phrase Scope, 16 participants accessed phrase definitions and 12 participants viewed hypothetical scenarios. Out of 77 instances when Phrase Scope was accessed, 72 were suggested by TermSight. This demonstrates the value of suggesting difficult or ambiguous phrases as opposed to solely relying on users to notice them.

\begin{figure*}[!ht]
    \centering
    \includegraphics[width=0.9\textwidth, height=0.8\textheight]
    {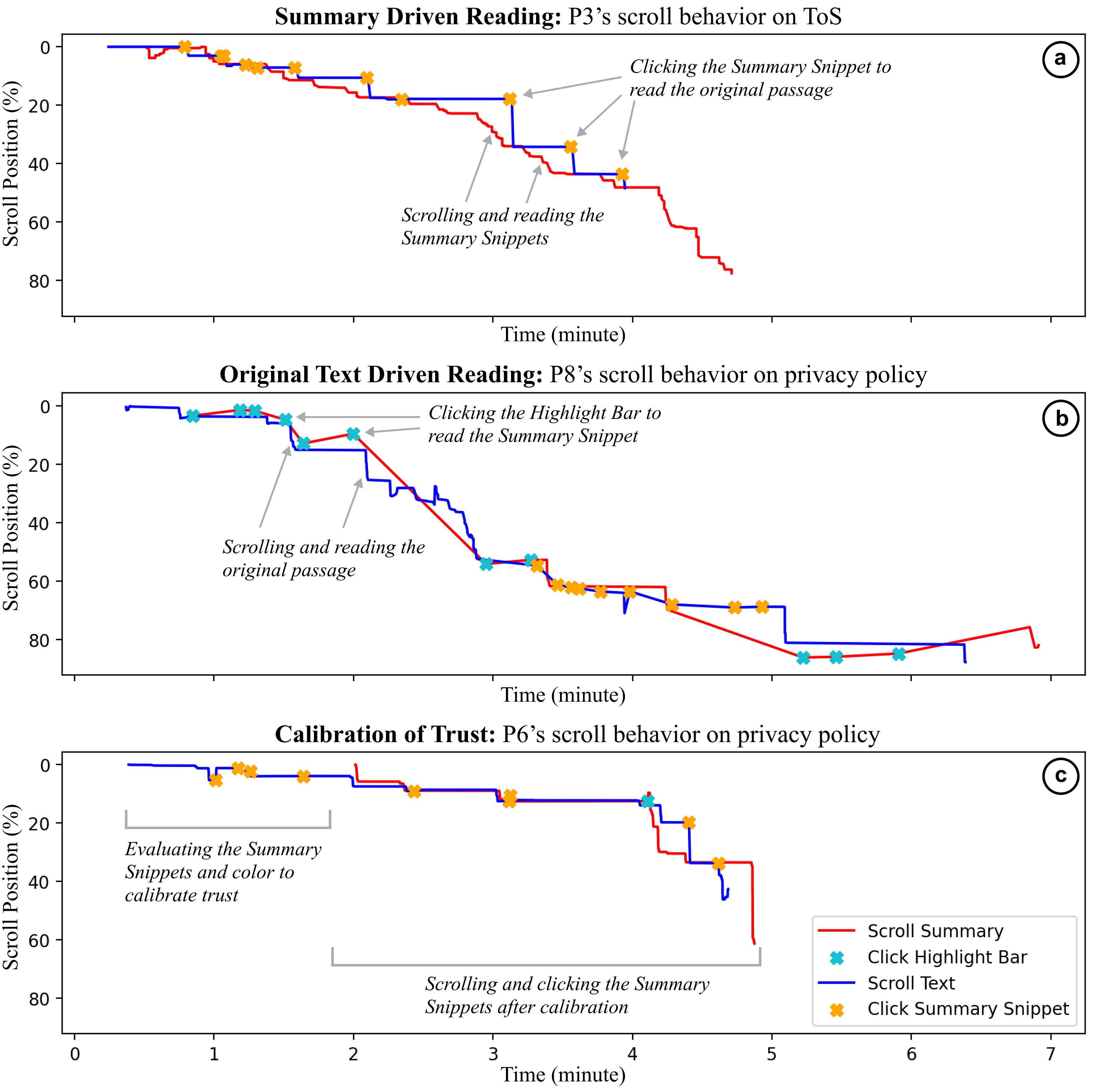}
    \caption{Scrolling behavior of participants who employed each non-exclusive reading strategy. TermSight features two scrollable columns: Summary Snippets on the left (red lines) and Original Text on the right (blue lines). Clicking a Summary Snippet (orange X) auto-scrolls the Original Text panel to the referenced text, while clicking the Highlight Bar (light blue X) auto-scrolls the Summary panel to the corresponding Summary Snippet. (a) Summary-driven reading occurs when participants primarily read and scroll the Summary Snippets, occasionally clicking on them to interrogate the original text (19/20). (b) Original text-driven reading occurs when participants first read the original text and use the Summary Snippets as a supplement (9/20). (c) The calibration of trust mainly occurs at the beginning of the session, where participants would compare and contrast the original text to the Summary Snippets with color (6/20).}
    \label{fig:readingstrategy}
    \Description{Visualization of the scrolling behavior of participants who employed each of the three non-exclusive reading strategies. (a) Summary-driven reading occurs when participants primarily read and scroll the Summary Snippets, occasionally clicking on them to interrogate the original text (19/20). (b) Original text-driven reading occurs when participants first read the original text and use the Summary Snippets as a supplement (9/20). (c) The calibration of trust mainly occurs at the beginning of the session where participants would compare and contrast the original text to the Summary Snippets with color (6/20).}
\end{figure*}

\subsubsection{Reading Strategies for Accessing Original Text}
\label{subsec:readingStrategies}

We observed diverse ways in which participants used TermSight to read ToS (Figure ~\ref{fig:readingstrategy}). Interestingly, these reading strategies centered around the use of the original text. In some cases, participants used TermSight's features as \textit{entries into} the original text, while in others, the features were used as \textit{verification of} the original text. In still others, participants used features in tandem with the text to \textit{calibrate trust} in TermSight. Below, we provide case studies of these strategies.

\paragraph{Summary driven reading}
19 out of 20 participants at times relied on the Summary Snippets as a primary entry point for engaging with the text. One notable example of this behavior was P3, who mainly read the Summary Snippets and clicked 11 of them to interrogate the original text (Figure ~\ref{fig:readingstrategy}a). P3 described how colors were driving his attention to read. P3's decision to click the Summary Snippets and interrogate the original text relied on the color coding and evaluation of the information loss of the Summary Snippets. 

\begin{quote}
    \textit{``I mainly stay on the left side. I read them if it was like a saturated red or green. I clicked on them because I felt like there was more information for me [behind the Summary], and I want to learn more about it.''} (P3)
\end{quote}

\paragraph{Original text driven reading}
On the other hand, nine participants went through at least one policy focusing mainly on the original text. 
These participants described how they wanted to first read the actual text before relying on AI-generated summaries due to worries about the potential imperfections of AI. 
When reading the privacy policy, P8 focused on the original text as shown by the frequent scrolling of the original text in Figure ~\ref{fig:readingstrategy}b. In addition, P8 clicked the Highlight Bar 11 times to refer back to the Summary Snippets. 
P8 described how he knew AI was not perfect from his prior professional experiences in AI. As a result, P8 took the colors and Summary Snippets as a guide rather than a guarantee.
\begin{quote}
    \textit{``I'm not gonna just go and trust it without doing my due diligence. I would take it as a guide. I won't take it as a guarantee, because AI is not perfect. I think it [AI] would try to label it the best it can and I think it did, to be honest.''} (P8)
\end{quote}

\paragraph{Calibration of trust}
Six participants described that they went through a calibration process with AI-generated features such as the Summary Snippets and color coding. For example, P6 described how his reading behavior was different during the calibration process and after. At the beginning of the reading session, P6 focused on reading the original text to evaluate the colors and summaries  (original text driven reading). Afterward, P6 relied on the summaries a lot more by scrolling and skimming the left panel and clicking on the Summary Snippets to dive to the original text, as seen in Figure ~\ref{fig:readingstrategy}c (summary driven reading).

\begin{quote}
    \textit{``Early on in the reading, I read the whole thing on the right first. Then, I looked at the summaries, and once I realized that it did do a good job summarizing it with colors. I trusted it enough for the rest of the time.''} (P6)
\end{quote}

\section{Discussion}

In this paper, we explored how AI-powered systems might support ToS contract reading given the unique challenges that legal text poses (i.e., value-dependent, ambiguous, and legally binding). Our formative study revealed the barriers ToS presents, which informed the design of an augmented reading interface, TermSight. Results from our user study (N=20) suggest that TermSight's features made ToS contracts easier to read and navigate, without an increase or decrease in comprehension. Below, we discuss the implications of our findings around the barriers ToS contracts present (\S\ref{subsec:formdis}), the design implications for future contract reading interfaces (\S\ref{sec:designimplications}), and the limitations of technology in mitigating the societal difficulties with service contracts (\S\ref{sec:techlimitation}).

\subsection{From Simplifying Legal Text to Navigating Dark Patterns}
\label{subsec:formdis}

Past work on augmenting legal text has focused on simplification (e.g., ~\cite{manor2019plain}). In the context of ToS contracts, our formative study revealed that the barriers people face extend well beyond complex text. Participants struggled to navigate multiple policies within a ToS, surface relevant information from dense legal language, and interpret ambiguous terminology. Surprisingly, existing navigational affordances---such as policy names, section headers, or overview summaries---often obfuscated rather than clarified the original text, a design reminiscent of dark patterns in the HCI literature ~\cite{mathur2021makes, bosch2016tales, kyi2023investigating}.
These barriers may not be unique to ToS, but reflect broader challenges for reading legal contracts taken to their extreme. For example, leases, insurance policies, and employment contracts similarly feature multiple addenda; long, dense text; and ambiguous language that obscures consumer rights ~\cite{mueller1970residential, furth2017unexpected}.
Future work could expand our findings by conducting large-scale analyses of dark patterns in contracts, an area largely underexplored by the HCI community. 
In addition, future contract reading interfaces should consider going beyond simplifying legal text to helping users navigate contracts while combating the dark patterns of design. TermSight serves as an initial exploration for inspiring future contract reading tools in this direction.

\subsection{Design Implications and Opportunities}
\label{sec:designimplications}

TermSight is a preliminary exploration of designing contract reading interfaces. Below, we discuss the design implications of our findings and opportunities for further exploration.

\subsubsection{\textbf{Providing guidance at multiple levels of granularity}} 

While prior work focuses on simplifying legal text (e.g., ~\cite{manor2019plain}),
TermSight demonstrated the potential values of guiding readers at the contract-level (e.g., Power Meter and document previews), document-level (e.g., color-coded Summary Snippets), and phrase-level (e.g., phrase identification). Participants in our user study affirmed these designs in both self-report and interaction behaviors (\S\ref{sec:Findings}). 
For example, participants reported how Power Meter transformed deciding which policy to read from a ``guessing game'' to a more deliberate decision-making process. 
On the other hand, Summary Snippets with color and Phrase Scope drew participants' attention to snippets of information and phrases buried in dense text that they might otherwise miss.
Consequently, finding which policy to read and what text to read within a policy was significantly easier for participants with TermSight.
The specificity of TermSight's guidance (i.e., relevance classification) relies on the user persona. For the user study, we designed two personas to be able to control for their quality (\S\ref{sec:userstudy_materials}). These personas also aligned with participants' profiles in the user study (Figure ~\ref{fig:userstudyparticipant}). Future work could explore allowing users to further customize their personas to receive personalized guidance and overcome the barriers of doing so (e.g., forming information bubbles ~\cite{sharma2024generative}).

\subsubsection{\textbf{Prompting reflection on unintended consequences}}

Prior work has explored providing term definitions to aid reading academic papers~\cite{august2023paper, head2021augmenting}. However, legal contracts are unique because contracts are agreements for the unforeseen future rather than a description of the past ~\cite{hart2017incomplete}. 
Reflecting this characteristic of contract, participants in our formative study mentioned that definitions alone may not always be sufficient for legal language because it can be difficult to contextualize abstract phrases in unforeseen future scenarios. 
In the user study, participants described the value of Phrase Scope for presenting scenarios to prompt reflection on possible implications of signing a contract, similar to how prior work has encouraged scientists to reflect on the unintended consequences of their research~\cite{wang2024farsight, pang2024blip}. Some participants even noted how the scenario was more helpful than the definition (\S\ref{subsubsec:phrasescopevalue}). 
Future work could further explore how to design these scenarios and investigate their influence on user perception of the contractual clauses, such as by manipulating the linguistic framing~\cite{sharma2023cognitive} or showing information from legal cases and news articles.

\subsubsection{\textbf{Centering augmentations around the original text}}

Legal contracts differ fundamentally from documents like academic papers or news articles because the language itself carries legal power \citep{chen2012contract}. 
Consequently, LM-powered text transformations cannot substitute the original text. TermSight explored the approach of tightly linking generated text with the original text through visual placement (i.e, side-by-side views) and deep linking (i.e., linking to narrowly attributed text like a single clause), allowing readers to directly compare language and navigate to the original text in at most one click. 
When given these features, participants took advantage of them to engage with the original text (\S\ref{subsec:readingStrategies}). Some participants dove into the original text by finding relevant Summary Snippets, while others verified their understanding of the original text with the Summary Snippets. Some participants even explicitly evaluated the Summary Snippets with the original text to calibrate trust in the system. 
An exciting avenue for future work would be to go beyond supporting original text reading to encouraging it. Features that nudge readers toward the original text where appropriate, such as revealing summaries only after reading ~\cite{chen2023marvista}, are an interesting start, but little work has evaluated these features' effects in situations where the original text might be discouraging to engage with.

\subsection{Technological Limitations and Legal Implications}
\label{sec:techlimitation}

Compared to simplified reading tasks (e.g., reading a single clause~\cite{robinson2020beyond, lee2025could} or policy overview~\cite{tabassum2018increasing, korunovska2020challenges}), our user study involved realistic ToS with over 10 policies.
TermSight made ToS significantly easier to read and navigate. However, TermSight did not worsen or improve comprehension, reminiscent of prior findings that simplified language often did not improve comprehension of privacy policies ~\cite{tabassum2018increasing, korunovska2020challenges, robinson2020beyond}.
There may be multiple explanations behind this finding. For example, Summary Snippets may have distracted attention away from the original text, the 12-word limit of Summary Snippets may have constrained the conceptual coverage of the original text, and participants' calibration behaviors (i.e., confirming summary information with the original text) may have distracted them from the reading task. 
However, our findings may have also revealed the broader limitations of technological solutions in mitigating the societal difficulties with service contracts such as ToS.
Although TermSight offered multi-level guidance on relevance (e.g., Power Meter, Summary Snippets), the sheer volume of potentially relevant information in ToS remains a dominant challenge.
For example, in the social media ToS, 172/759 (23\%) discrete Information Snippets were classified by TermSight as being relevant to the given user persona (\S\ref{sec:userstudy_materials}), including 6 snippets needed to answer the comprehension questions.
Within this subset, being able to notice and recall snippets of information to correctly answer the comprehension questions may be difficult. 
While future deployment of TermSight could allow users to customize the personas for more targeted guidance, this may risk introducing information bubbles 
~\cite{adamic2005political,conover2011political, garrett2009echo, sharma2024generative}. 
Given that the formation of a binding contract under U.S. law is premised on the doctrine of the `duty to read'\footnote{Contracting parties are legally responsible for reading and understanding the contract before signing~\cite{benoliel2019duty}.}, our findings raise fundamental questions about the practicality of this duty for ToS. Instead of relying solely on technological fixes, policy and legal innovations are needed. For example, policy makers and legal scholars could explore alternative paradigms of contracting that operationalize ToS into shorter, context-specific contracts to reduce the upfront burden to consent.

\subsection{Limitations and Future Work}
\label{sec:limitations}
The nature of this work is exploratory. Similar to prior HCI studies (e.g., ~\cite{chen2023marvista, cao2023dataparticles, kambhamettu2024explainable}), we designed and studied TermSight to explore the challenges and opportunities of designing contract reading interfaces, rather than proposing TermSight as the definitive solution. Future work can pursue finer-grained exploration through controlled ablation studies. Examples may include investigating the influence of Power Meter on ToS reading frequency in the wild, assessing how hypothetical scenarios affect user perception of clauses (e.g., trust ~\cite{shneiderman2000designing}), or investigating how AI affordances (e.g., Summary Snippets) support or distract from original text reading (e.g, \cite{nielsen2023effects}) through eye-tracking.

Because participants were paid to read ToS in a timed reading task, their motivation and behavior may not reflect real-world usage. Moreover, a within-subject design may lead to ordering effects or fatigue. In the study, we counterbalanced the conditions and didn't observe a significant effect of ordering on user experience, comprehension, and recall (Appendix Figure ~\ref{fig:forest_ordering}). Yet, it may still have influences on participants' interaction. Future work can expand our study by conducting between-subject or field experiments. 

In addition, our participants were recruited from Prolific, mostly college-educated, and fluent in English, which may limit the generalization of our findings to the broader population. Marginalized communities, including neurodivergent readers and individuals with limited English literacy, can face additional barriers when interpreting legal documents. 
For example, readers with ADHD might have a hard time staying focused when reading visually dense text such as ToS ~\cite{stern2013role, bental2007relationship}. Future work can investigate how to help alternative reader populations make sense of legal contracts by evaluating TermSight with these populations of users or integrating additional supports (e.g., translation). 

Future systems could adapt our design of TermSight to other contracts. For example, Summary Snippets may help readers surface and interpret relevant or predatory clauses in leases, while Phrase Scope may help patients envision the unintended consequences of signing a medical contract through patient-specific scenarios.
Finally, features of TermSight rely on the capabilities of LMs, which may produce imperfect outputs ~\cite{huang2024survey, maynez2020faithfulness, Luo2024Factual}. While work is actively investigating how to guide LMs to generate factually correct information ~\cite{Vu2023FreshLLMs}, we believe a validation mechanism for LLM output is necessary for future deployment of systems like TermSight, such as by facilitating end-user~\cite{lam2022end, jahanbakhsh2022our} or expert auditing of AI output~\cite{jiang2024leveraging}.

\section{Conclusion}

We regularly sign legal contracts for where we live, who we work with, and how we interact with digital services. Yet, these contracts have drifted from their premise of facilitating mutual understanding to complex documents that discourage reading.
In this work, we explored the opportunities and limitations of intelligent reading support for legal contracts, using ToS contracts as a case study. 
Our formative study revealed ineffective and deceptive designs at all levels of a ToS. 
To make service contracts approachable, we designed and evaluated an intelligent reading interface: TermSight. Participants reported that TermSight reduced the difficulty of engaging with ToS and increased their willingness to do so.
Additionally, features of TermSight enabled participants the ability to surface and approach relevant information at all levels of contract reading. 
Taken together, TermSight presents one avenue for making legal contracts more approachable to the general public. However, facilitating a deeper understanding of ToS remains an open challenge due to the overwhelming amount of information in ToS.

\bibliographystyle{ACM-Reference-Format}
\bibliography{manuscript}

\appendix

\section{Iterative Design Study}
\label{appendix_iterativedesign}

A total of 8 participants were recruited through Prolific to evaluate an early prototype of TermSight. This prototype closely resembled the current version of TermSight but lacked a document preview when hovering over the Power Meter. Additionally, the prototype used three saturation levels for relevance, resulting in a total of nine colors (3 hues for power x 3 saturation levels for relevance). Participants could also toggle between two layouts for the Summary Snippets: condensed layout and in-context layout.
The condensed layout matched the current design shown in Figure ~\ref{fig:termsight}. In contrast, the in-context layout placed each Summary Snippet directly next to its corresponding Information Snippet.
After a brief interface tutorial, participants were given 10 minutes to read a ToS using the interface, followed by a semi-structured interview about their experience. Each session lasted for 40 minutes. Below, we present the main findings and changes made.

\paragraph{\textbf{Participants preferred condensed layout for quick navigation}}
7 participants preferred and primarily used the condensed layout of the Summary Snippets during the reading session. They noted that the purpose of the Summary Snippets was to support navigation, and the condensed layout made it easier to gain an overview and identify relevant information in a policy.
In contrast, the in-context layout required significantly more scrolling, which participants found overwhelming. As a result, in the final version of TermSight, we used the condensed layout (Figure ~\ref{fig:termsight}). 

\paragraph{\textbf{Reduce visual complexity of the color scheme}}
All participants found the colors to be intuitive and effective in highlighting power and relevance, helping them decide what to read. However, 3 participants pointed out that while the power dimension was intuitive, the relevance dimension with 3 saturation levels was challenging to differentiate, as there would be 9 colors in total. To reduce visual complexity, the final version of TermSight included only two saturation levels (High vs. Low) for relevance (Figure ~\ref{fig:colorscheme}).

\paragraph{\textbf{Complement Power Meter with Document Preview}}
Additionally, 4 participants suggested that while the Power Meter provided a general overview, they wanted a more concrete preview. This feedback led to the integration of a document preview feature in the final version of TermSight, which appears when users hover over the Power Meter (Figure ~\ref{fig:powermeter}).

\section{Additional Implementation Details}
\label{Appendix:implementationdetails}


\subsection{Source Document and Pre-processing}
\label{Appendix:docpreprocessing}

Given a source file in HTML or markdown, the document is first segmented by headers (e.g., h1, h2, h3, h4). Within each section or subsection, the text is further chunked by newline separators (i.e., "\textbackslash n") into segments of around 1,500 characters (approximately 250 words) using langchain's RecursiveCharacterTextSplitter\footnote{https://python.langchain.com/v0.1/docs/modules/\linebreak data\_connection/document\_transformers/recursive\_text\_splitter/}, with no overlap between chunks. Importantly, paragraph structures are preserved, as the text splitter only splits at newline breaks.

\subsection{Obtaining Summary Snippets and Information Snippets}
\label{Appendix:obtainsummary}

The prompt used to obtain the Summary Snippets and Information Snippets is detailed in Figure ~\ref{fig:Prompt_summarysnippet}. We constrained the Summary Snippets to be short because prior works have found that adding short summaries (10-20 words) under search results was more effective for navigational and information-seeking tasks compared to longer summaries, which can be harder to skim ~\cite{cutrell2007you, sweeney2006effective}. 

\subsection{Classifying Information Snippets}
\label{Appendix:classifyinfos}

In TermSight, two classifications were performed for each Information Snippet using GPT-4o with few-shot prompting (Figure ~\ref{fig:Prompt-classify-power} and ~\ref{fig:Prompt-classify-rel}). For the classification of Power, each Information Snippet was classified by the degree of control or agency it grants to the Service Provider or the User (Categories: Service, Neutral, User). For the classification of Relevance, Information Snippets were classified based on their relevance to the user persona (Categories: High, Low). User personas used for the study are included in Appendix ~\ref{Appendix_persona}.

\subsection{Phrase Scope}
\label{Appendix:phrasescope}

\subsubsection{Identifying unfamiliar or ambiguous phrases}
\label{Appendix:identifyphrase}

For each document chunk, we employ two few-shot prompts to GPT-4o to identify potentially unfamiliar (Figure ~\ref{fig:Prompt-identifyphraseunfamiliar}) or vague phrases (Figure ~\ref{fig:Prompt-identifyphrasevague}). 
Few-shot examples for identifying unfamiliar and ambiguous phrases were selected based on phrases participants found challenging in our formative study.
Both prompts were applied to each document chunk, producing two sets of identified phrases. Since a phrase may be both unfamiliar and vague, the union of these two sets was taken as the final set of identified phrases for a given chunk of text.

\subsubsection{Generating phrase definitions and answers to user questions}
\label{Appendix:gendefinitionanswer}

We first retrieve potentially relevant document chunks from the vector database. The database query is framed as a question: \textit{``What does \{input phrase\} refer to in the sentence: \{phrase context\}''}. Here, phrase context refers to the chunk of text containing the input phrase. Then, the query is embedded using OpenAI's text-embedding-3-small model, and the top 15 chunks from the vector database are retrieved based on cosine similarity. Figure ~\ref{fig:Prompt-def} shows the prompt used to generate the definitions after retrieving potentially relevant document chunks. When users ask additional questions, the same retrieval-augmented question answering pipeline used for generating definitions is applied, with the only difference being the question asked (Figure ~\ref{fig:Prompt-ask}).

\subsubsection{Generating scenarios}

We leveraged GPT-4o to generate customized scenarios of potential implications based on user persona. The prompt is specified in Figure ~\ref{fig:Prompt-scenario}.

\begin{table*}[!ht]
\centering
\small
\renewcommand{\arraystretch}{1.05}
\begin{tabular}{p{2cm} p{5cm} p{1cm} p{5cm}}
\hline
\textbf{Classification} & \textbf{Input Snippet} & \textbf{Output} & \textbf{Imperfection} \\
\hline
Power & 
``All Buy Now purchases in a ServiceY Show are final and binding.'' & 
Service & 
Lack of enough context in the input. Users do have the option to return. It's a neutral clause. \\

Power & 
``Use public content for any illegal, deceptive, unethical, false, misleading, or improper purpose, including the infringement of third-party intellectual property rights.'' & 
Neutral & 
Lack of enough context in the input. The clause describes what third party licensee cannot do with user content. It's a user-benefiting clause. \\

Relevance & 
``Your earnings are based on the listing price and actual earnings will vary based on the final order price, Seller discounts, and any other applicable taxes and discounts.'' & 
High & 
Lack of enough context in the input. This clause is for sellers and is less relevant to the input buyer persona. \\

Relevance & 
``ServiceY cannot guarantee that a ServiceY consignment listing will be sold or that a certain sales amount will be earned for individual items or an entire shipment.'' & 
High &
Lack of enough context in the input. This clause is for sellers and is less relevant to the input buyer persona. \\
\hline
\end{tabular}
\caption{Examples of imperfect classifications of Power and Relevance.}
\label{table:imperfect-class}
\Description{Examples of imperfect classifications of Power and Relevance. The four columns describe the type of classification (Relevance or Power), model input, model output, and the imperfection.}
\end{table*}

\section{Evaluation of TermSight Output}
\label{appendix_system_eval}

\subsection{Classification of Information Snippets}

Out of the sampled 116 snippets, our evaluation revealed 3 instances where the power classifications were imperfect, mainly caused by the lack of context in the input Information Snippet. For example, the snippet (\textit{`All Buy Now purchases in a ServiceY Show are final and binding'}) was classified by TermSight as favoring the service. In the generated explanation, the assumption was that the statement meant that the user can not return or refund the purchase. However, users can return and get a refund if there are problems with the purchase (e.g., an item doesn't match its description) as stated in the service's return policy, making this snippet more neutral. The return policy was not included in the snippet context; as a result, this neutral statement was considered more service-oriented. We didn’t observe any clause favoring the service provider being misclassified as neutral or favoring the user.

For the classification of relevance, there were 2 instances, out of 116, where irrelevant Information Snippets were classified as relevant to the persona. In both cases, the snippets were more relevant to sellers, not buyers, even though the persona given was that of a buyer. 
There were 11 instances where the Information Snippets classified as relevant were indirectly relevant to the input user personas. For example, sellers have to pay a fee to the platform after selling an item. Despite this information being more relevant to sellers, TermSight classified it as relevant to buyers, as it exposed the potential hidden fees that sellers might include as part of their listing price. We show examples of imperfect classification of Power and Relevance in Table ~\ref{table:imperfect-class}.

\subsection{Term Definitions and Scenarios}

Our evaluation revealed that all the generated definitions were correct, but out of the 113 definitions, 4 were only general definitions of the phrase, not specific to the ToS. On further inspection, we identified that this was because these vague phrases were not explicitly defined anywhere in the ToS. Additionally, their meanings are service-specific and cannot be extrapolated from common sense or inferred by large language models (LLMs). For example, for the phrase \textit{`aggregated anonymized statistics'}, TermSight provided a general definition of what it might mean to aggregate and anonymize user data. However, the specific details---such as what user data is being aggregated---were not specified in the ToS. Rather than a limitation of TermSight, these imperfections highlight the ill-defined nature of the language used in ToS. 

For the generated scenarios of phrases relevant to the user persona, we found one instance where the scenario was factually incorrect based on the input context. The input context and phrase stated that users do not gain ownership rights by downloading content from the service. However, the scenario claimed that users might lose ownership rights over the content they create by uploading it to the service. We also noticed that when the passages or phrases target a different audience (e.g., developers) than the user persona used to generate the scenarios (i.e., a lay user), the generated scenarios become less relevant or useful.
We did not regenerate these imperfections to keep the user experience realistic in real-world settings where the LLM outputs are not guaranteed to be perfect. Participants in our user study were informed that AI output can be imperfect. This specific incorrect scenario was not accessed by any participants in the study.

\section{Study Materials: Formative Study}
\label{Appendix_prelim}

Questions asked before the reading session:
\begin{itemize}
    \item How familiar are you with Terms of Service in general?
    \item Have you previously read or wished that you read the Terms of Service? What were your reasons for wanting or not wanting to read the Terms of Service? 
    \item Have you used your assigned service before?
    \item Have you read the ToS for your assigned service before?
\end{itemize}
Semi-structured interview questions asked after the reading session:
\begin{itemize}
    \item What were the challenges you faced when reading the Terms of Service?
    \item How did you go about reading the ToS?
    \item What information were you interested in in ToS? Both from your prior experience in reading ToS and this ToS?
    \item Imagine if you have a magic wand that can transform the ToS in whatever ways you want. How would you transform the Terms of Service?
\end{itemize}

\section{Study Materials: User Study}
\label{Appendix_comparativestudy}

\subsection{Baseline Interface}
\begin{figure*}[!ht]
    \centering
    \includegraphics[width=0.9\textwidth]{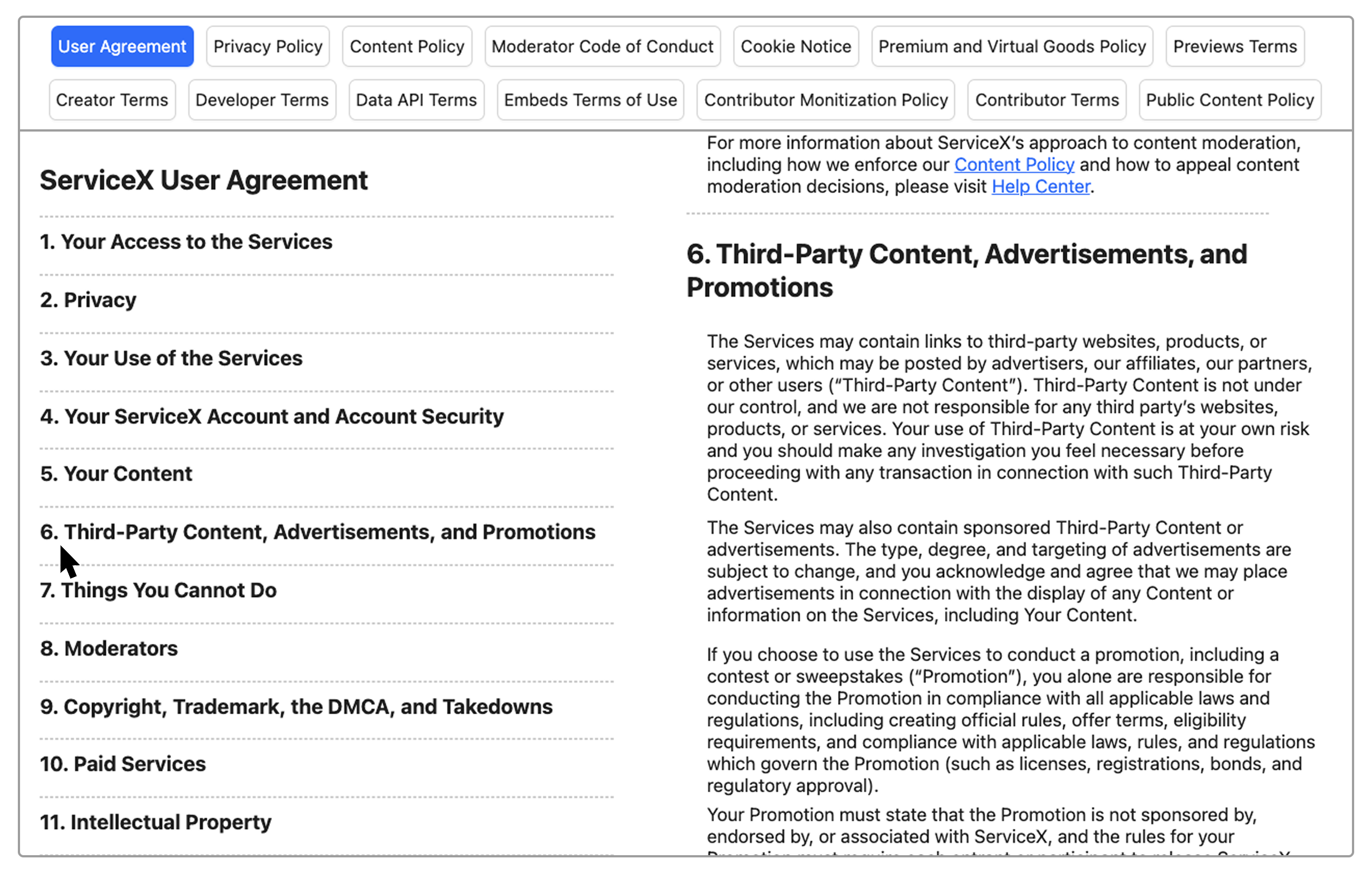}
    \caption{The baseline reading interface. Participants can navigate to different policies using the top navigational panel. In place of the Summary Snippets on the left is a table of contents. Participants can click a section header in the table of contents to navigate to the corresponding section.}
    \label{fig:termsightbaseline}
    \Description{The baseline system used in the user study. The interface contains three panels: a top navigation panel for switching between different policies, the original text on the right, and the table of contents on the left. Participants could click a section header in the table of contents to navigate to the corresponding section.}
\end{figure*}

The baseline ToS reading interface used in the user study can be found in Figure ~\ref{fig:termsightbaseline}.

\subsection{User Persona}
\label{Appendix_persona}

Two personas were given to the participants during the user study for the social media service (content consumer who posts personal content) and the e-commerce service (buyer who rarely posts reviews). The same personas were used for features of TermSight to classify relevance and generate personalized scenarios. The personas were designed based on information that participants in the formative study described caring about.

\aptLtoX[graphic=no,type=html]{\begin{framed}
\small
\textbf{Persona: Content consumer who posts personal content}    

Imagine you are a lay user of social media platforms. You are over 18 years old and located in the United States. 

    Your usage of Social Media sites: 
    \begin{itemize}
    \item You spend most of your time on the platform scrolling through feeds, liking posts, chatting with other users, and sharing personal content such as photos.
    \end{itemize}
    
    Things you care about when using Social Media Sites: 
    \begin{itemize}
    \item You care about Privacy, particularly what data is being collected and how your data can be used and shared.
    \item You care about what the service can do with user-generated content, such as licenses over user content or advertising with user content.
    \item You care about potential liabilities when using ServiceX.
    \end{itemize}
\end{framed}}{\begin{center}
\fcolorbox{gray!60}{white}{
\begin{minipage}{0.92\linewidth}
\small
\textbf{Persona: Content consumer who posts personal content}    

Imagine you are a lay user of social media platforms. You are over 18 years old and located in the United States. 

    Your usage of Social Media sites: 
    \begin{itemize}
    \item You spend most of your time on the platform scrolling through feeds, liking posts, chatting with other users, and sharing personal content such as photos.
    \end{itemize}
    
    Things you care about when using Social Media Sites: 
    \begin{itemize}
    \item You care about Privacy, particularly what data is being collected and how your data can be used and shared.
    \item You care about what the service can do with user-generated content, such as licenses over user content or advertising with user content.
    \item You care about potential liabilities when using ServiceX.
    \end{itemize}
\end{minipage}
}
\end{center}}

\aptLtoX[graphic=no,type=html]{\begin{framed}
\small
\textbf{Persona: Buyer who rarely posts reviews}    

    Imagine you are a lay user of E-commerce platforms. You are over 18 years old and located in the United States. 

    Your usage of E-commerce sites:
    \begin{itemize}
    \item You typically engage with the E-commerce platform to buy new or used items from other users.
    \item You rarely post any reviews or content on the service.
    \end{itemize}
    
    Things you care about when using E-commerce sites:
    \begin{itemize}
    \item You care about information related to making purchases, refunds, returns, user protection policies, termination, arbitration, and liabilities.
    \item You also care about Privacy, particularly what data is being collected and how your data can be used and shared.
    \end{itemize}
\end{framed}}{\begin{center}
\fcolorbox{gray!60}{white}{
\begin{minipage}{0.92\linewidth}
\small
\textbf{Persona: Buyer who rarely posts reviews}    

    Imagine you are a lay user of E-commerce platforms. You are over 18 years old and located in the United States. 

    Your usage of E-commerce sites:
    \begin{itemize}
    \item You typically engage with the E-commerce platform to buy new or used items from other users.
    \item You rarely post any reviews or content on the service.
    \end{itemize}
    
    Things you care about when using E-commerce sites:
    \begin{itemize}
    \item You care about information related to making purchases, refunds, returns, user protection policies, termination, arbitration, and liabilities.
    \item You also care about Privacy, particularly what data is being collected and how your data can be used and shared.
    \end{itemize}
\end{minipage}
}
\end{center}}

\subsection{Participant Demographics}
\label{Appendix_demographics}
\begin{figure}[!ht]
    \centering
    \includegraphics[width=0.45\textwidth]{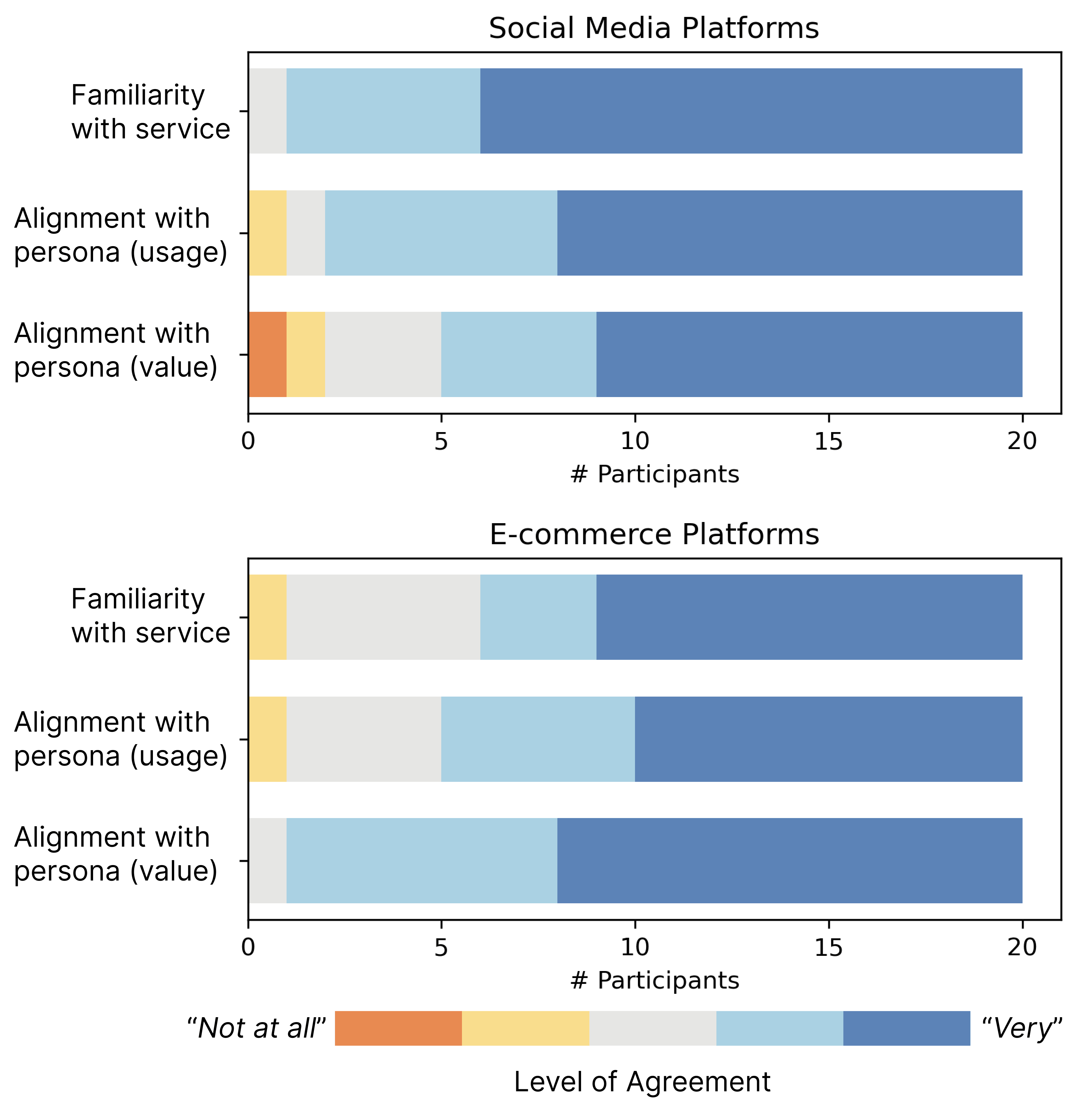}
    \caption{Participants' self-ratings of their familiarity with social media and e-commerce platforms and their alignment with the user persona given for each service. Most participants were familiar with both social media and e-commerce platforms and found the given persona to align with their personal usage of the service and personal value.}
    \label{fig:userstudyparticipant}
    \Description{Participants' self-ratings of their general familiarity with social media and e-commerce platforms and their alignment with the user persona given for each service. Most participants were familiar with both social media and e-commerce platforms in general and found the given persona to align with their personal usage of the service and personal value.}
\end{figure}

Of the 20 participants, 13 self-identify as female and 7 as male. 1 participant had a high school degree, 3 had an associate degree, 12 had a bachelor's degree, and 4 had a master's degree. 4 participants were between the ages of 18-25, 5 were between 26-35, 5 were between 36-45, and 6 were between 46-55. When asked about how many ToS they've read, 4 participants read none, 5 read between 1--3, 5 read between 4--6, 1 read between 7--9, and 5 read greater than 10. Participants in the user study also found their given user persona for both services to highly align with their personal usage of the service and their personal value (Figure ~\ref{fig:userstudyparticipant}).

\subsection{Pre-survey Questions}
\label{Appendix_pre-survey}

Questions that were asked once at the beginning of the interview: 
\begin{itemize}
    \item For how many online platforms have you read their Terms of Service (ToS) before? [None (0), Few (1-3), Some (4-6), Many (7-9), A lot (>10)]
    \item How familiar are you with Terms of Service (ToS) for online platforms? (5-point Likert rating)
\end{itemize}
5-point Likert rating questions that were asked before each of the two reading sessions:
\begin{itemize}
    \item How familiar are you with (E-commerce or Social Media) sites?

    \item How well does the above user persona align with your personal usage of (E-commerce or Social Media) sites?

    \item How well does the above user persona align with things you personally care about when using (E-commerce or Social Media) sites?
\end{itemize}

\subsection{5-point Likert Ratings of Reading Experience}
\label{Appendix_likert}

Participants rated their reading experience after each of the two reading sessions.

\begin{itemize}
    \item \textbf{E1:} I’m interested in spending more time reading the service's Terms of Service (ToS) with the current interface and wish to get a link after the study.
    \item \textbf{E2:} How hard did you have to work to read the ToS?
    \item \textbf{E3:} How easy was it for you to decide which sub-policies to read?
    \item \textbf{E4:} How easy was it for you to decide what text to read within a sub-policy?
    \item \textbf{E5:} How confident are you that you got all the relevant information from the ToS (including sub-policies)?
    \item \textbf{E6:} How much do you feel like you understood the ToS?
    \item \textbf{E7:} How much would you be willing to read the ToS of other services with this interface?
\end{itemize}

\subsection{Semi-structured Interview Questions}
\label{Appendix_interviewQ}

Below are questions that were asked after each of the two reading sessions, except the last question. The last question was only asked after the second reading session, where participants would compare and contrast both interfaces.  
\begin{itemize}
    \item Describe your experience reading the ToS using the interface.
    \item How did you read the ToS using the interface? 
    \item What were the challenges you faced when reading the Terms of Service? To what extent did the interface help you? (Be specific about the features in the interface). 
    \item Which interface do you prefer? Why? Compare and contrast. (Asked only after the second reading session)
\end{itemize}

\section{Bayesian Analysis Model Details and Additional Analysis}
\label{sec:Model Details}

\begin{figure*}[!ht]
    \centering
    \includegraphics[width=0.75\textwidth]{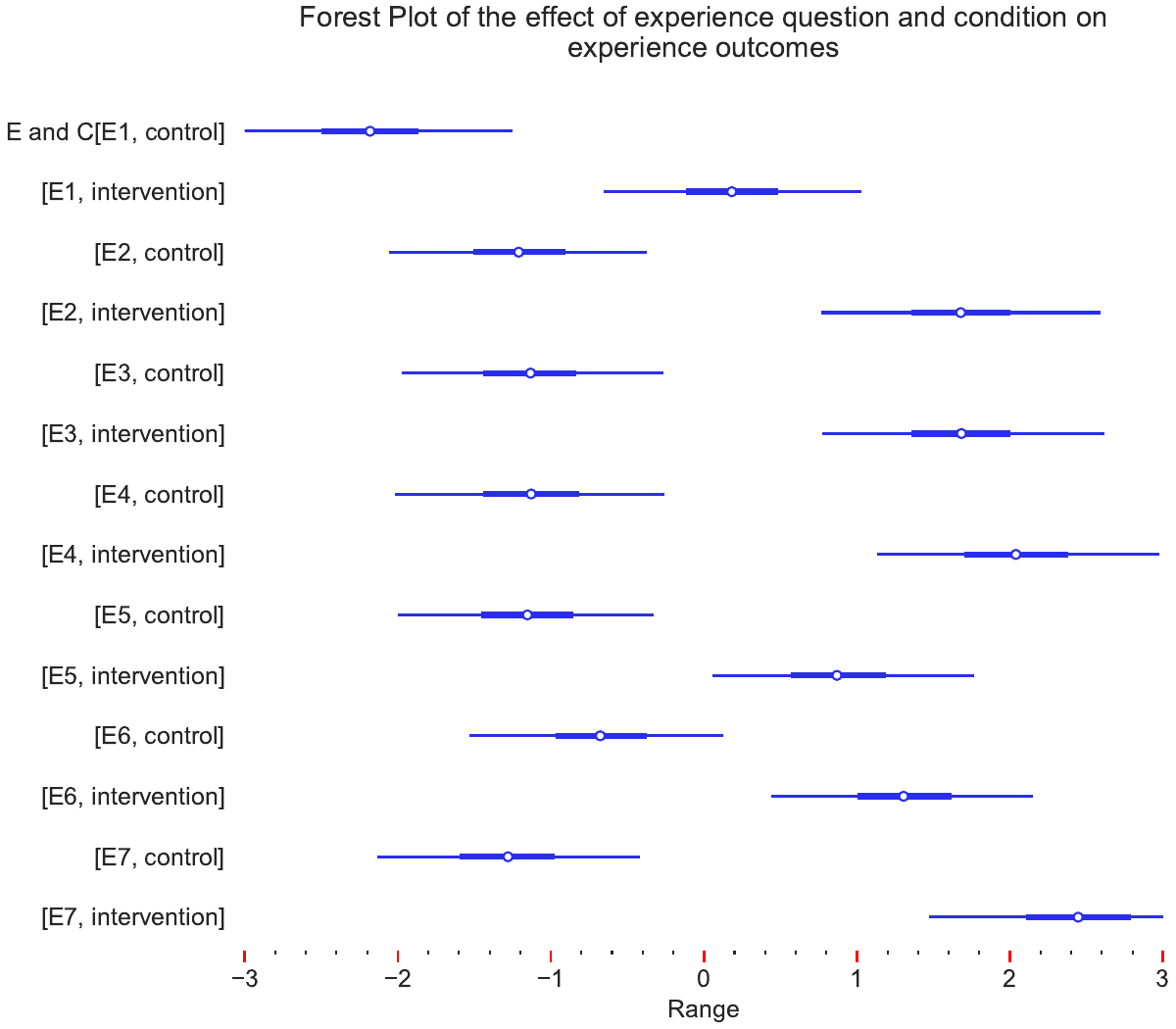}
    \caption{The figure shows a forest plot of coefficients in the model, corresponding to treatment and control, for each of the experience questions. Each line of the plot shows a 94\% High Density Interval (HPDI) for the coefficient. The inner, thicker line represents the 50\% HPDI. The results show a significant effect of the TermSight interface on the user experience for every question. Since the 94\% HPDI for each pair (control, treatment) for every question does not overlap with each other, we should expect a significant effect of the treatment on the user experience. E1--E7 refer to the 7 experience questions specified in Appendix \ref{Appendix_likert}.}
    \label{fig:forest_comprehension}
    \Description{The figure shows a forest plot of coefficients in the model, corresponding to treatment and control, for each of the 7 experience questions. Each line of the plot shows a 94\% High Density Interval (HPDI) for the coefficient. The inner, thicker line represents the 50\% HPDI. There are a total of 14 lines (7 pairs of 2) in the plot. The results show a significant effect of the TermSight interface on the user experience for every question. Since the 94\% HPDI for each pair (control, treatment) for every question does not overlap with each other, we should expect a significant effect of the treatment on the user experience. The 7 experience questions are specified in Appendix E.5.}
\end{figure*}

\begin{figure*}[!ht]
    \centering
    \includegraphics[height=0.8\textheight]{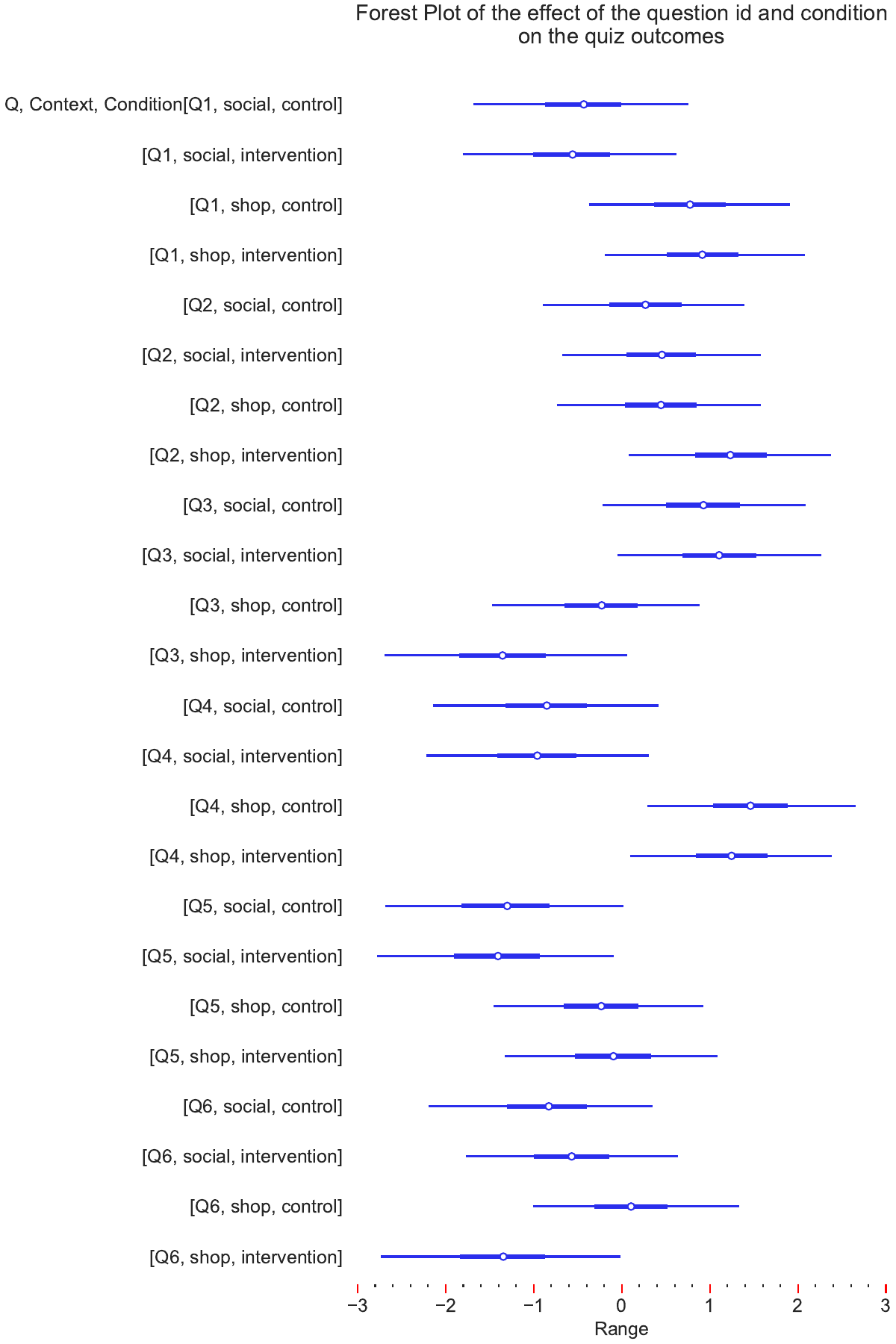}
    \caption{The figure shows a forest plot of coefficients in the model, corresponding to treatment and control, for each of the comprehension questions. The $x$-axis is on the logistic scale, with +3 corresponding to a 0.95 probability value on the outcome scale (-3 corresponds to 0.05). Each line of the plot shows a 94\% High Density Interval (HPDI) for the coefficient. The inner, thicker line represents the 50\% HPDI. The 94\% HPDI intervals for each treatment--control pair, when the question and service type are fixed, show significant overlap. This suggests no significant effect of the treatment on the user's comprehension. Note that the comprehension questions Q1--Q6 are different for the two service types and are included in the supplemental materials.}
    \label{fig:forest_quiz}
    \Description{The figure shows a forest plot of coefficients in the model, corresponding to treatment and control, for each of the comprehension questions. There are a total of 24 lines (12 pairs of 2). The x-axis is on the logistic scale, with +3 corresponding to a 0.95 probability value on the outcome scale (-3 corresponds to 0.05). Each line of the plot shows a 94\% High Density Interval (HPDI) for the coefficient. The inner, thicker line represents the 50\% HPDI. The 94\% HPDI intervals for each treatment–control pair, when the question and service type are fixed, show significant overlap. This suggests no significant effect of the treatment on the user’s comprehension for every question. Note that the comprehension questions Q1–Q6 are different for the two service types and are included in the supplemental materials.}
\end{figure*}

\begin{figure*}[!ht]
    \centering
    \includegraphics[width=0.75\textwidth]{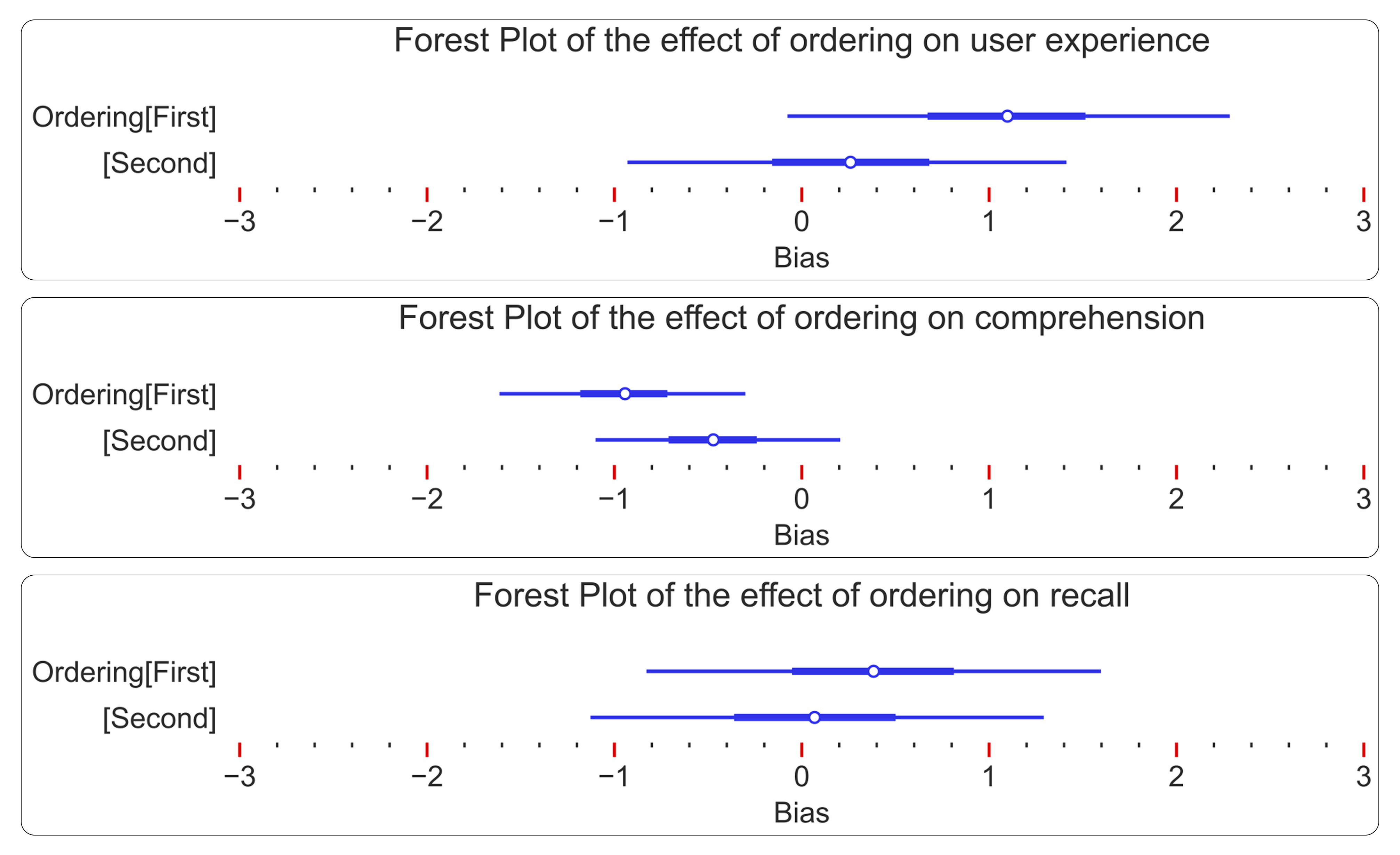}
    \caption{The figure shows the forest plots of coefficients in the model for user experience, comprehension, and recall with respect to the ordering. Each line of the plot shows a 94\% High Density Interval (HPDI) for the coefficient. The inner, thicker line represents the 50\% HPDI. Since the 94\% HPDI for each ordering pair (first, second) overlaps, there is no significant effect of ordering on user experience, comprehension, and recall.}
    \label{fig:forest_ordering}
    \Description{The figure shows the forest plots of coefficients in the model for user experience, comprehension, and recall with respect to the ordering. Each line of the plot shows a 94\% High Density Interval (HPDI) for the coefficient. The inner, thicker line represents the 50\% HPDI. Since the 94\% HPDI for each ordering pair (first, second) overlaps, there is no significant effect of ordering for user experience, comprehension, and recall.}
\end{figure*}

In this section, we provide details on the Bayesian models. 
We first explain the concepts of structural causal framework and DAG in \S\ref{sub:Structural Causal Framework}.
All models were informed by the DAG in Figure ~\ref{fig:dags}. 
Next, in \S\ref{sub:Modeling Experience Outcomes}, we describe the model used to estimate the effect of the treatment on the user experience. In \S\ref{sub:Modeling Comprehension Outcomes}, we describe the model used to estimate the effect of the treatment on the comprehension outcomes, i.e., the quiz questions. Finally, in \S\ref{sub:Modeling Recall Ouctomes}, we describe the model used to estimate the effect of the treatment on the experiment subjects' ability to recall facts correctly. 
Figure ~\ref{fig:forest_comprehension} and ~\ref{fig:forest_quiz} show the forest plots of model coefficients for the experience and comprehension outcomes. 
Figure ~\ref{fig:forest_ordering} presents forest plots illustrating the effect of ordering on user experience, comprehension, and recall.

\subsection{Structural Causal Framework}
\label{sub:Structural Causal Framework}
We used a structural causal framework popularized by~\citet{Pearl2009} to better understand the effects of the treatment. A structural causal model framework comprises exogenous variables $U$, endogenous variables $V$, and a set of functions $F$ that represent the physical process that results in the values of the endogenous variables. For example, the question $v_i = f(v, u)$ implies that the values of $v$ and $u$ completely determine the value of $v_i$ through the function $f$. The structural causal model can be represented as a Directed Acyclic Graph (DAG) where the nodes represent the variables and the edges represent the causal relationships among the variables. Thus, in the preceding example, $v$ and $u$ are the parents of $v_i$ in the DAG. The structural causal model framework is non-parametric, but frequently used with generalized linear models to estimate the functions $f$. The structural causal framework has connections with structural equation models~\cite{Wright1934} used in the Economic Sciences, though the latter were developed with linear relationships in mind. Figure ~\ref{fig:dags} shows the DAG for the study.

\subsection{Modeling Experience Outcomes}
\label{sub:Modeling Experience Outcomes}

We asked the experiment subjects to rate their experience of using the treatment or the baseline interface on a Likert scale from 1 to 5 (L=5 outcome values). There were N subjects, and we asked each subject K (K=7) questions. Since the Likert scale is ordinal, we modeled the experience outcomes using an ordinal regression model. The DAG in Figure ~\ref{fig:dags} indicates that the experimental condition alone affects the experience outcomes. There were three randomizations: whether the subject was shown the treatment or control interface (i.e., interface type), whether the subject was shown the social media ToS or the e-commerce ToS (i.e., service type), and whether the treatment was shown first or second (i.e., presentation order). We modeled the experience outcomes using a cumulative \texttt{logit} model with a linear link function:

\begin{align}
    Y_{i,j,k,l,m} &\sim \mathrm{OrderedLogistic}(\kappa, \phi_{i,j,k,l,m}) \quad \\
    \phi_{i,j,k,l, m} &= \alpha_i + c_{j,k} +  s_{l} + o_{m} \quad 
\end{align}

Below, we specify the priors for $\alpha_{i}$ (intercept, one per participant), $c_{j,k}$ (joint variable for the interface type and user experience item), $s_{l}$ (service type), $o_{m}$ (presentation order), and $\kappa_i$ (cutpoints).

\begin{align}
    \alpha_i &\sim \mathrm{Normal}(0, 1), i \in \{1, \dots,N\} \quad \\
    c_{j,k} &\sim \mathrm{Normal}(0, 1), j \in \{1,2\}, k \in \{1, \dots, K\}\quad \\
    s_{l} &\sim \mathrm{Normal}(0, 1), l \in \{1,2\} \quad  \\
    o_{m} &\sim \mathrm{Normal}(0, 1), m \in \{1,2\} \quad \\
    \kappa_i &\sim \mathrm{Normal}(0, 1), i \in \{1, \dots, L-1\} \quad \label{eq:cutpoints}
\end{align}

In eq (\ref{eq:cutpoints}), we further ensured that the samples are ordered, i.e., $\kappa_1 < \kappa_2 < \dots < \kappa_{L-1}$. The priors are conservative. For example, on the logit scale, a coefficient that is normally distributed with mean 0 and standard deviation 1, the range [-3, 3] covers 99\% of the distribution and evaluates using inverse logistic to the outcome probability range of [0.04, 0.95]. 

\subsection{Modeling Comprehension Outcomes}
\label{sub:Modeling Comprehension Outcomes}

There were N subjects, and we asked each subject K (K=6) questions to test their comprehension. The outcome is a binary variable with 1 indicating a correct answer and 0 indicating an incorrect answer. The DAG in Figure ~\ref{fig:dags} indicates that the experimental condition alone affects the comprehension outcomes. There were three randomizations: whether the subject was shown the treatment or control interface, whether the subject was shown the social media ToS or the e-commerce ToS, and whether the treatment was shown first or second. We modeled the comprehension outcomes using a logistic regression model. Notice that we created a joint variable $c_{j,k,l}$ for the coefficients corresponding to the interface type, question, and service type. We did this because the comprehension questions were different across the two services used in the experiment. We included the interface variable (i.e., treatment vs. control) to be able to easily estimate the effect of the treatment on the comprehension outcomes per question. The model is given by:

\begin{align}
    Y_{i,j,k,l,m} &\sim \mathrm{Binomial}(p_{i,j,k,l,m}) \quad \\
    \mathrm{Logistic}(p_{i,j,k,l,m}) &= \alpha_i + c_{j,k,l}  + o_{m} \quad  
\end{align}

Below, we specify the priors for $\alpha_{i}$ (intercept, one per participant), $c_{j,k,l}$ (joint variable for the interface type, comprehension question, and service type), and $o_{m}$ (presentation order). 

\begin{align}
    \alpha_i &\sim \mathrm{Normal}(0, 1), i \in \{1, \dots,N\} \quad \\
    c_{j,k,l} &\sim \mathrm{Normal}(0, 1), j \in \{1,2\}, k \in \{1, \dots, K\}, l \in \{1,2\}\quad  \\
    o_{m} &\sim \mathrm{Normal}(0, 1), m \in \{1,2\} \quad 
\end{align}

As in the previous model, the priors are conservative. For example, on the logit scale, a coefficient that is normally distributed with mean 0 and standard deviation 1, the range [-3, 3] covers 99\% of the distribution and evaluates using inverse logistic to the outcome probability range of [0.04, 0.95].

\subsection{Modeling Recall Outcomes}
\label{sub:Modeling Recall Ouctomes}

There were N subjects, and we asked each subject to recall facts from the ToS. The outcome that we measure is the number of correctly recalled facts, making the outcome a count variable. Since many of the subjects could not recall any facts correctly, we modeled the recall outcomes using a Zero Inflated Poisson regression model. The DAG in Figure ~\ref{fig:dags} indicates that the experimental condition alone affects the comprehension outcomes. As before, there were three randomizations: whether the subject was shown the treatment or control interface, whether the subject was shown the social media ToS or the shopping ToS, and whether the treatment was shown first or second. 
The model is given by:

\begin{align}
    Y_{i,j,k,l} &\sim \mathrm{ZeroInflatedPoisson}(\lambda_{i,j,k,l}, \phi) \quad  \\
    \mathrm{Log}(\lambda_{i,j,k,l}) &= \alpha_i + c_{j} +  s_{k} + o_{l} \quad  
\end{align}

Below, we specify the priors for $\alpha_{i}$ (intercept, one per participant), $c_{j}$ (interface type), $s_{k}$ (service type), $o_{l}$ (presentation order), and $\phi$ (probability of zero inflation). 

\begin{align}
    \alpha_i &\sim \mathrm{Normal}(0, 1), i \in \{1, \dots,N\} \quad \\
    c_{j} &\sim \mathrm{Normal}(0, 1), j \in \{1,2\}\quad  \\
    s_{k} &\sim \mathrm{Normal}(0, 1), k \in \{1,2\} \quad  \\
    o_{l} &\sim \mathrm{Normal}(0, 1), l \in \{1,2\} \quad  \\
    \phi &\sim \mathrm{Beta}(2, 2) \quad 
\end{align}

As in the previous model, the priors are conservative. For example, on the Log scale, a coefficient that is normally distributed with mean 0 and standard deviation 1, the range [-3, 3] covers 99\% of the distribution and evaluates using inverse log to the outcome range of [0.05, 21]. Notice that we model the number of correctly recalled facts, and thus is a conservative prior. Given that most subjects could not recall any facts, the zero inflation parameter is set to a weakly informative prior using a Beta distribution.

\section{Supplemental Analysis for Comprehension Outcomes}
\label{sub:Additional_comp_analysis}

We aimed to design the comprehension questions to have a single answer that best matches the clauses in the original text. However, we acknowledge that legal problems and contractual clauses may leave room for interpretation ~\cite{marotta2025building}. As a result, we run an additional, more conservative analysis by removing any question that may contain room for interpretation on whether the original text would apply to the situation being asked in the question, leading to different answer choices. We removed 2 out of 6 questions for each service type (Social Media: Q1, Q4; Shopping: Q2, Q3). For all other questions, there is a single correct answer choice directly matching the original text, while all the other options either conflict with or were never mentioned in the original text. Below is one example of the questions removed from this analysis:

\setlength\fboxsep{8pt} 
\setlength\fboxrule{0.5pt} 

\aptLtoX[graphic=no,type=html]{\begin{framed}
\small
\textbf{Social Media Q1:} How can ServiceX use photographs you post for advertisements or promotions?\newline

\textbf{Original Answer:} ServiceX can use photos you post in ads without your permission and is not obligated to attribute you as the creator.\newline

\textbf{Relevant Clause:} ``...you grant us a worldwide, royalty-free, perpetual, irrevocable, non-exclusive, transferable, and sublicensable license to use, copy, modify, adapt, prepare derivative works of, distribute, store, perform, and display Your Content and any name, username, voice, or likeness provided in connection with Your Content in all media formats and channels now known or later developed anywhere in the world...you irrevocably waive any claims and assertions of moral rights or attribution with respect to Your Content.'' (\textit{User Agreement})\newline

\textbf{Possible Room for Interpretation:} Though the original answer best matches the relevant clause. The clause does not explicitly mention advertisement. This may leave room for interpretation on whether the clause applies to ads.
\end{framed}}{\begin{center}
\fcolorbox{gray!60}{white}{
\begin{minipage}{0.9\linewidth}
\small
\textbf{Social Media Q1:} How can ServiceX use photographs you post for advertisements or promotions?\\[0.3em]
\textbf{Original Answer:} ServiceX can use photos you post in ads without your permission and is not obligated to attribute you as the creator.\\[0.3em]
\textbf{Relevant Clause:} ``...you grant us a worldwide, royalty-free, perpetual, irrevocable, non-exclusive, transferable, and sublicensable license to use, copy, modify, adapt, prepare derivative works of, distribute, store, perform, and display Your Content and any name, username, voice, or likeness provided in connection with Your Content in all media formats and channels now known or later developed anywhere in the world...you irrevocably waive any claims and assertions of moral rights or attribution with respect to Your Content.'' (\textit{User Agreement})\\[0.3em]
\textbf{Possible Room for Interpretation:} Though the original answer best matches the relevant clause. The clause does not explicitly mention advertisement. This may leave room for interpretation on whether the clause applies to ads.
\end{minipage}
}
\end{center}}

We use the same Bayesian model as \S\ref{sub:Modeling Comprehension Outcomes}, except that there are 4 comprehension questions for each service that are analyzed (i.e., K=4).
Out of the 4 questions for each service, participants on average scored 1.5 (\textit{\(\sigma=1.0\)}) when using TermSight and 1.6 (\textit{\(\sigma=1.1\)}) when using the baseline interface. Our Bayesian model analysis reveals no significant differences in the comprehension scores across interface conditions. The posterior distribution of the contrast between treatment and control on comprehension is centered around zero and overlaps with ROPE. When the question ID and service type are fixed, there are significant overlaps between the 94\% HPDI for each treatment--control pair, suggesting no significant differences for every question. These findings matches with our original findings. The figures for this additional analysis can be found in the supplemental materials (e.g., forest plots).

\begin{figure*}[t]
\centering
\small

\fcolorbox{black!30}{gray!25}{%
  \begin{minipage}{0.95\linewidth}
  \textbf{Prompt for Obtaining Summary Snippets and Information Snippets}
  \end{minipage}%
}

\vspace{-0.2em}

\fcolorbox{black!25}{gray!5}{%
  \begin{minipage}{0.95\linewidth}
  \small

Summarize the input section of the Terms of Service into concise bullet points (less than 12 words) in plain language. When adjacent paragraphs or sentences share a similar or related theme, only output 1 single bullet point. For each bullet point summary, include the full-text reference to the original passage in \{\} and don't use "..." to reduce text in the reference. When outputting a reference, don't change anything in the original text, such as spaces and newlines. There can be multiple sentences or paragraphs that reference a single summary.  The references to summary should cover the original text.\\

Example output format: \textbf{\{EXAMPLE OUTPUT\}} \\

Input: \textbf{\{INPUT TEXT CHUNK\}} \\
    
  \end{minipage}%
}

    \caption{Prompt for obtaining Summary Snippets and Information Snippets.}
    \label{fig:Prompt_summarysnippet}
    \Description{Prompt for obtaining Summary Snippets and Information Snippets. 
Summarize the input section of the terms of service into concise bullet points (less than 12 words) in plain language. When adjacent paragraphs or sentences share a similar or related theme, only output 1 single bullet point. For each bullet point summary, include the full-text reference to the original passage in {} and don’t use "..." to reduce text in the reference. When outputting a reference, don’t change anything in the original text, such as spaces and newlines. There can be multiple sentences or paragraphs that reference a single summary. The references to the summary should
cover the original text. Example output format: {EXAMPLE OUTPUT} Input: {INPUT TEXT CHUNK}.
}
\end{figure*}

\begin{figure*}[t]
\centering
\small

\fcolorbox{black!30}{gray!25}{%
  \begin{minipage}{0.95\linewidth}
  \textbf{Prompt for Classifying Power}
  \end{minipage}%
}

\vspace{-0.2em}

\fcolorbox{black!25}{gray!5}{%
  \begin{minipage}{0.95\linewidth}
  \small

           Classify the input term from a Terms of Service agreement based on the power relationship and benefit between the service and the user. Use the following categories (Service, Neutral, User):\newline
          - Service: The term grants the service provider disproportionate power or control over the user. It may impose unfair restrictions, obligations, or liabilities on the user, or reduce the user's rights and autonomy over their data or content.\newline
          - Neutral: The term outlines standard procedures, responsibilities, or conditions the user and service have. For example, users take responsibility for the content they post. It neither significantly favors the service provider nor the user, and does not substantially impact the user's rights.\newline
          - User: The term empowers the user by offering clear protections, rights, or benefits, ensuring transparency, and limiting the service provider's power.\newline
          
          Examples for each category:
          
          Service:\newline
          - The service can delete specific content without prior notice and without a reason.\newline
          - The service can license user content to third parties.\newline
          - The service tracks your personal data for advertising
          
          Neutral:\newline
          - Users are responsible for the content they post\newline
          - Users agree not to use the service for illegal purposes\newline
          - Blocking first-party cookies may limit your ability to use the service
          
          User:\newline
          - You can opt out of targeted advertising\newline
          - The service does not sell your personal data\newline
          - The service will not allow third parties to access your personal information without a legal basis\newline
          
          Output format in JSON:
          \{"Category": "Service/Neutral/User",
          "Explanation": "explanation of output" \}\newline

          Input: \textbf{\{INPUT INFORMATION SNIPPET\}}
    
  \end{minipage}%
}

    \caption{Prompt for classifying the power balance of the Information Snippets.}
    \label{fig:Prompt-classify-power}
    \Description{Prompt for classifying the power balance of the Information Snippets. 
Classify the input term from a Terms of Service agreement based on the power relationship and benefit between the service and the user. Use the following categories (Service, Neutral, User):
 - Service: The term grants the service provider disproportionate power or control over the user. It may impose unfair restrictions, obligations, or liabilities on the user, or reduce the user's rights and autonomy over their data or content.
 - Neutral: The term outlines standard procedures, responsibilities, or conditions the user and service have. For example, users take responsibility for the content they post. It neither significantly favors the service provider nor the user, and does not substantially impact the user's rights.
 - User: The term empowers the user by offering clear protections, rights, or benefits, ensuring transparency, and limiting the service provider's power.
Examples for each category:
Service:
- The service can delete specific content without prior notice and without a reason.\newline
- The service can license user content to third parties.
- The service tracks your personal data for advertising
Neutral:
- Users are responsible for the content they post
- Users agree not to use the service for illegal purposes
- Blocking first party cookies may limit your ability to use the service
User:
- You can opt out of targeted advertising
- The service does not sell your personal data
- The service will not allow third parties to access your personal information without a legal basis
 Output format in JSON: 
{"Category": "Service/Neutral/User",
“Explanation": "explanation of output" }
Input: {INPUT INFORMATION SNIPPET}
}
\end{figure*}

\begin{figure*}[t]
\centering
\small

\fcolorbox{black!30}{gray!25}{%
  \begin{minipage}{0.95\linewidth}
  \textbf{Prompt for Classifying Relevance}
  \end{minipage}%
}

\vspace{-0.2em}

\fcolorbox{black!25}{gray!5}{%
  \begin{minipage}{0.95\linewidth}
  \small

    For the input term from a Terms of Service, output a relevance rating (High/Low) of the input term with respect to the user persona.\newline
    [High]: The term is directly relevant to the user's usage of the service or what the user cares about. The term applies to the user persona and is necessary for the user to know.\newline
    [Low]: The term is not relevant to the user's usage of the service or what the user cares about. The term doesn't apply to the user persona or is not necessary for the user to know.\newline
    
    User Persona: \textbf{\{INPUT USER PERSONA\}} \newline

    Output format in JSON: \{"Relevance": "Low/High", "Explanation": "explanation of output" \}\newline

    Input: \textbf{\{INPUT INFORMATION SNIPPET\}}
    
  \end{minipage}%
}

    \caption{Prompt for classifying the relevance of Information Snippets to the user persona.}
    \label{fig:Prompt-classify-rel}
    \Description{Prompt for classifying the relevance of Information Snippets to the user persona.
For the input term from a Terms of Service, assign a relevance rating (High/Low) of the input term with respect to the user persona.
Relevance rating:
[High]: The term is directly relevant to the user’s usage of the service or what the user cares
about. The term applies to the user persona and is necessary for the user to know.
[Low]: The term is not relevant to the user’s usage of the service or what the user cares about. The term doesn’t apply to the user persona or is not necessary for the user to know.
User Persona: {INPUT USER PERSONA}
Output format in JSON: {"Relevance": "Low/High", "Explanation": "explanation of output" }
Input: {INPUT INFORMATION SNIPPET}
}
\end{figure*}

\begin{figure*}[t]
\centering
\small

\fcolorbox{black!30}{gray!25}{%
  \begin{minipage}{0.95\linewidth}
  \textbf{Prompt for Identifying Unfamiliar Phrases}
  \end{minipage}%
}

\vspace{-0.2em}

\fcolorbox{black!25}{gray!5}{%
  \begin{minipage}{0.95\linewidth}
  \small

          You are a helpful assistant who extracts words or multi-word phrases in the input section of Terms of Service that a high schooler might not know the meaning of. Jargon refers to domain-specific terminologies that a lay user might not know about.\newline
          
          Example jargon: \newline
          - legal jargon: indemnity, arbitration, liability\newline
          - copyright licenses: sublicensable licenses, royalty-free licenses\newline
          - technical privacy terms: cookies, Ad identifiers, Authentication tokens\newline
          
          Return an empty array if the section does not contain jargon. The extracted word should exactly match the original input text with the same capitalization and sequence of words. \newline
          
          Output format in JSON: \{"Jargon": []\}\newline

          Input: \textbf{\{INPUT TEXT CHUNK\}}
    
  \end{minipage}%
}

    \caption{Prompt for identifying potentially unfamiliar phrases for lay users in a text chunk.}
    \label{fig:Prompt-identifyphraseunfamiliar}
    \Description{Prompt for identifying potentially unfamiliar phrases for lay users in a text chunk.
You are a helpful assistant who extracts words or multi-word phrases in the input section of Terms of Service that a high schooler might not know the meaning of. Jargon refers to domain-specific terminologies that a lay user might not know about.
Example jargon:
- legal jargon: indemnity, arbitration, liability
- copyright licenses: sublicensable licenses, royalty-free licenses
- technical privacy terms: cookies, Ad identifiers, Authentication tokens
Return an empty array if the section does not contain jargon. The extracted word should exactly match the original input text with the same capitalization and sequence of words.
Output format in JSON: {"Jargon": []}
Input: {INPUT TEXT CHUNK}
}
\end{figure*}

\begin{figure*}[t]
\centering
\small

\fcolorbox{black!30}{gray!25}{%
  \begin{minipage}{0.95\linewidth}
  \textbf{Prompt for Identifying Vague Phrases}
  \end{minipage}%
}

\vspace{-0.2em}

\fcolorbox{black!25}{gray!5}{%
  \begin{minipage}{0.95\linewidth}
  \small

          You are a helpful legal assistant who extracts vague terms (can have multiple words in one term) in the input section of Terms of Service. A vague term refers to information that is vaguely abstracted without a clear definition provided in the section. \newline
          
          Example Vague terms: information, other, some, third parties, most, generally, personal data,  others, general, many, various, might, services, certain information\newline
          
          Return an empty array if the section does not contain vague terms. The extracted word should exactly match the original input text with the same capitalization and sequence of words.\newline
          
          Output format in JSON: \{"Vague": []\}\newline

          Input: \textbf{\{INPUT TEXT CHUNK\}}
    
  \end{minipage}%
}

    \caption{Prompt for identifying ambiguous phrases in a text chunk.}
    \label{fig:Prompt-identifyphrasevague}
    \Description{Prompt for identifying ambiguous phrases in a text chunk.
You are a helpful legal assistant who extracts vague terms (can have multiple words in one term) in the input section of Terms of Service. A vague term refers to information that is vaguely
abstracted without a clear definition provided in the section.
Example Vague terms: information, other, some, third parties, most, generally, personal data,
others, general, many, various, might, services, certain information
Return an empty array if the section does not contain vague terms. The extracted word should
exactly match the original input text with the same capitalization and sequence of words.
Output format in JSON: {"Vague": []}
Input: {INPUT TEXT CHUNK}
}
\end{figure*}

\begin{figure*}[t]
\centering
\small

\fcolorbox{black!30}{gray!25}{%
  \begin{minipage}{0.95\linewidth}
  \textbf{Prompt for Generating Phrase Definitions}
  \end{minipage}%
}

\vspace{-0.2em}

\fcolorbox{black!25}{gray!5}{%
  \begin{minipage}{0.95\linewidth}
  \small

    Use information in the retrieved context to provide a definition of the user-selected phrase or term. Avoid using long sentences. For example, if the user-selected term is "information", define what the term "information" includes and refers to, such as: location data, interaction data, profile data, etc. The output definition should be specific and straight to the point; don't include language that doesn't contribute to the definition, such as 'in the given context'. Output the string list of reference ids (["ref1", ...]) used to generate the definition under "References". If the definition of the phrase is not specified in the retrieved context, output a definition of what the phrase might mean and output an empty array for "References".\newline
    
    Examples: \textbf{\{EXAMPLES\}} \newline
    
    Output format in JSON: \{"Definition": "", "References": ["ref1", "ref2", "ref3"]\} \newline
    
    Retrieved Context: \textbf{\{RETRIEVED CONTEXT\}} \newline

    Question: What does \textbf{\{INPUT PHRASE\}} refer to?\newline
    Context around the user-selected phrase: \textbf{\{PHRASE CONTEXT\}}\newline
    
  \end{minipage}%
}

    \caption{Prompt for generating in-context phrase definitions.}
    \label{fig:Prompt-def}
    \Description{Prompt for generating in context phrase definitions.
Use information in the retrieved context to provide a definition of the user-selected phrase
or term. Avoid using long sentences. For example, if the user-selected term is "information",
define what the term "information" includes and refers to, such as: location data, interaction
data, profile data, etc. The output definition should be specific and straight to the point;
don’t include language that doesn’t contribute to the definition, such as ’in the given context’.
Output the string list of reference ids (["ref1", ...]) used to generate the definition under
"References". If the definition of the phrase is not specified in the retrieved context, output a
definition of what the phrase might mean and output an empty array for "References".
Examples: {EXAMPLES}
Output format in JSON: {"Definition": "", "References": ["ref1", "ref2", "ref3"]}
Retrieved Context: {RETRIEVED CONTEXT}
Question: What does {INPUT PHRASE} refer to?
Context around the user-selected phrase: {PHRASE CONTEXT}
}
\end{figure*}

\begin{figure*}[t]
\centering
\small

\fcolorbox{black!30}{gray!25}{%
  \begin{minipage}{0.95\linewidth}
  \textbf{Prompt for Generating Scenarios}
  \end{minipage}%
}

\vspace{-0.2em}

\fcolorbox{black!25}{gray!5}{%
  \begin{minipage}{0.95\linewidth}
  \small

    Tell a concise what-if scenario or example in less than 50 words to demonstrate the meaning and potential implications of the user-selected phrase based on the context around the user-selected phrase. The scenario/example should be relevant to the below user persona using \{an E-commerce platform of used items / a Social Media platform\}. \newline
    
    User Persona: \textbf{\{INPUT USER PERSONA\}} \newline
    
    Output format in JSON: \{"Story": ""\}\newline
    
    User selected phrase: \textbf{\{INPUT PHRASE\}}\newline
    Context around the user-selected phrase: \textbf{\{PHRASE CONTEXT\}}\newline
    Definition of user selected phrase: \textbf{\{GENERATED DEFINITION\}}
    
  \end{minipage}%
}

    \caption{Prompt for generating scenarios to contextualize the meaning and potential implications of a phrase.}
    \label{fig:Prompt-scenario}
    \Description{Prompt for generating scenarios to contextualize the meaning and potential implications of a phrase.
Tell a concise what-if scenario or example in less than 50 words to demonstrate the meaning and potential implications of the user-selected phrase based on the context around the user-selected phrase. The scenario/example should be relevant to the below user persona using {an E-commerce platform of used items / a Social Media platform}.
User Persona: {INPUT USER PERSONA}
Output format in JSON: {"Story": ""}
User selected phrase: {INPUT PHRASE}
Context around the user-selected phrase: {PHRASE CONTEXT}
Definition of user selected phrase: {GENERATED DEFINITION}
}
\end{figure*}

\begin{figure*}[t]
\centering
\small

\fcolorbox{black!30}{gray!25}{%
  \begin{minipage}{0.95\linewidth}
  \textbf{Prompt for Generating Answers to User Questions.}
  \end{minipage}%
}

\vspace{-0.2em}

\fcolorbox{black!25}{gray!5}{%
  \begin{minipage}{0.95\linewidth}
  \small

    You are an assistant for question-answering tasks. Use information in the retrieved context to answer the user's question in less than 5 sentences. Output the string list of reference ids (["ref1", ...]) used to generate the definition under "References". If the definition of the phrase is not specified in the retrieved context, output a definition of what the phrase might mean and output an empty array for "References".\newline
    
    Examples: \textbf{\{EXAMPLES\}} \newline
    
    Output format in JSON: \{"Answer": "", "References": ["ref1", "ref2", "ref3"]\} \newline
    
    Retrieved Context: \textbf{\{RETRIEVED CONTEXT\}} \newline

    Question: \textbf{\{USER QUESTION\}}\newline
    User selected phrase: \textbf{\{INPUT PHRASE\}}\newline
    Context around the user-selected phrase: \textbf{\{PHRASE CONTEXT\}}
    
  \end{minipage}%
}

    \caption{Prompt for generating answers to user questions.}
    \label{fig:Prompt-ask}
    \Description{Prompt for generating answers to user questions.
You are an assistant for question-answering tasks. Use information in the retrieved context to
answer the user’s question in less than 5 sentences. Output the string list of reference ids
(["ref1", ...]) used to generate the definition under "References". If the definition of the
phrase is not specified in the retrieved context, output a definition of what the phrase might
mean and output an empty array for "References".
Examples: {EXAMPLES}
Output format in JSON: {"Answer": "", "References": ["ref1", "ref2", "ref3"]}
Retrieved Context: {RETRIEVED CONTEXT}
Question: {USER QUESTION}
User selected phrase: {INPUT PHRASE}
Context around the user-selected phrase: {PHRASE CONTEXT}
}
\end{figure*}

\begin{table*}[t]
\centering
\renewcommand{\arraystretch}{1.5} 
\begin{tabular}{p{1cm} p{5cm} p{4cm} p{5cm}}
\hline
\textbf{Phrase Type} & \textbf{Original Passage} & \textbf{Generated Definition}& \textbf{Generated Scenario} \\
\hline
Jargon & 
``... When Your Content is created with or submitted to the Services, you grant us a worldwide, royalty-free, perpetual, irrevocable, non-exclusive, transferable, and \textbf{**sublicensable**} license to use, copy, modify, adapt, prepare derivative works of, distribute, store, perform, and display Your Content and any name, username, voice, or likeness provided ...'' & 
Sublicensable refers to the ability to grant further licenses to third parties to use, copy, modify, or distribute your content. & 
Imagine posting a photo on ServiceX, and later discovering it's featured in a third-party app's advertisement. Due to the sublicensable license, ServiceX had the right to allow another company to use your photo, without needing your further permission, raising concerns about control over your content. \\

Jargon & 
``... The listed prices for Items do not include taxes, but the taxes will be displayed before a Buyer confirms the purchase. Use of \textbf{**ServiceY Credit**} (as defined below) may modify taxes that apply to a Buyer's purchase ... ''&
ServiceY Credit refers to non-redeemable promotional credits offered by ServiceY to be used exclusively for purchases on the Service. & 
Imagine Jane, a savvy shopper, receives \$10 ServiceY Credit for a promotion. She buys a vintage lamp listed at \$50. At checkout, ServiceY Credit reduces her total to \$40, modifying applicable taxes. Jane saves money, but can't withdraw or transfer the Credit—it only applies to ServiceY purchases. \\

Vague & 
``To use certain features of our Services, you may be required to create a ServiceX account (an Account id) and provide us with a username, password, and \textbf{**certain other information**} about yourself as set forth in the Privacy Policy ...'' & 
Certain other information refers to optional details such as a bio, gender, age, location, profile picture, or social link that you may provide when creating a ServiceX account. & 
Imagine signing up for ServiceX, and you're asked to provide a username, password, and 'certain other information' like your age and location. Later, you find out that ServiceX uses this data to tailor ads specifically for you, raising concerns about how much they know about you and potential privacy risks. \\

Vague & 
``... ServiceY reserves the right to discontinue providing Labels to any or all Users at any time and for \textbf{**any reason**} ...'' & 
Any reason refers to ServiceY's discretion to stop providing shipping labels without needing to specify a particular cause or justification. & 
Imagine you sell vintage clothes online. ServiceY provides you with prepaid shipping labels. Suddenly, without explanation, they stop offering these labels to you. This means you'll need to cover shipping costs yourself, impacting your profits. This demonstrates ServiceY's right to discontinue services for 'any reason,' affecting your business.\\

\hline
\end{tabular}
\caption{Examples of the generated definitions and scenarios for potentially unfamiliar and vague phrases.}
\label{table:example-def-scenario}
\Description{Four examples of the Phrase Scope’s generated definitions and scenarios for potentially unfamiliar and vague phrases. The four columns describe the type of phrase (Jargon or Vague phrase), the original passage that contains the phrase, the generated definition, and the generated scenario.}
\end{table*}

\end{document}